\documentclass[12pt,preprintnumbers,nofootinbib,letterpaper]{article}
\pdfoutput=1
\usepackage{amsmath,amssymb,amsfonts,cite,xcolor,epsf,epsfig,epstopdf,graphics,graphicx,mathtools,subcaption,physics,verbatim}
\def\thefootnote{*\arabic{footnote}}
\definecolor{ultramarine}{rgb}{0.07, 0.04, 0.56}
\definecolor{cadmiumgreen}{rgb}{0.0, 0.42, 0.24}
\definecolor{indigo(dye)}{rgb}{0.0, 0.25, 0.42}
\usepackage[linktocpage=true,breaklinks]{hyperref}
\hypersetup{
	colorlinks=true,
	citecolor=ultramarine,
	linkcolor=cadmiumgreen,
	urlcolor=indigo(dye),
}

\usepackage[utf8]{inputenc}

\usepackage{array}
\newcolumntype{P}[1]{>{\centering\arraybackslash}p{#1}}

\usepackage[shortlabels]{enumitem}

\usepackage{dcolumn}
\newcolumntype{M}[1]{>{\centering\arraybackslash}m{#1}}
\newcolumntype{N}{@{}m{0pt}@{}}

\usepackage{amsmath}
\usepackage{ulem}

\newcommand{\be}{\begin{equation}}  
\newcommand{\ee}{\end{equation}}

\usepackage{tikz}

\begin{document}


\begin{center}

\def\thefootnote{\fnsymbol{footnote}}

\vspace*{1.5cm}
{\Large {\bf Genesis--Starobinsky inflation can explain the ACT data}}
\\[1cm]

{Han Gil Choi$^{1}$, Pavel Petrov$^{1}$, and Seong Chan Park$^{2,3}$}
\\[.7cm]

{\small \textit{$^1$Cosmology, Gravity, and Astroparticle Physics Group, Center for Theoretical Physics of the Universe, Institute for Basic Science (IBS), Daejeon, 34126, Korea}}\\

{\small \textit{$^{2}$ Department of Physics and IPAP, Yonsei University,  Seoul, 03722, Korea}}\\

{\small \textit{$^{3}$ Korea Institute for Advanced Study, Seoul, 02455, Korea}}

\end{center}

\vspace{1.5cm}


\begin{abstract}

We propose a novel non-singular cosmological scenario within the framework of Horndeski gravity, consisting of three successive stages:
(i) a Genesis phase, in which the Universe slowly expands from an asymptotically flat spacetime;
(ii) a brief transition stage restoring General Relativity; and
(iii) a Starobinsky inflationary phase.
This construction is fully consistent within a viable parameter space: it remains weakly coupled, free from ghost and gradient instabilities, with luminal tensor and subluminal scalar perturbations throughout the entire evolution. Importantly, the Genesis phase induces characteristic corrections to the Starobinsky potential, which cannot be captured by simple $\sum_i c_i R^i$-type modifications. These corrections robustly enhance the scalar spectral index, thereby improving the agreement of Starobinsky inflation with recent CMB measurements, in particular the data from the Atacama Cosmology Telescope (ACT).

\end{abstract}
\vspace{0.5cm} 
\def\thefootnote{\arabic{footnote}}
\setcounter{footnote}{0}




\section{Introduction}
\label{sec: Introduction}

Nowadays, inflation is the prevailing paradigm for describing the primordial Universe. However, it faces  the problem of an initial singularity (see Ref.~\cite{Borde:1996pt}). This issue has motivated the exploration of non-singular cosmological scenarios, such as Genesis and bouncing Universe models. These alternatives aim to either replace or extend the standard inflationary framework. Realizing such scenarios typically requires violating  the Null Energy Condition (NEC), which in turn demands exotic forms of matter. A particularly promising approach involves Horndeski gravity~\cite{Horndeski:1974wa}, a well-motivated theoretical framework for constructing non-singular cosmologies.

Despite its promise, this approach encounters  significant challenges, most notably the so-called no-go theorem~\cite{Kobayashi:2016xpl, Libanov:2016kfc}, which states that any non-singular solution within Horndeski theory inevitably encounters gradient instabilities at some stage of its evolution. Two principal strategies have been proposed to overcome this obstacle. The first is to go beyond Horndeski theory, moving to more general frameworks such as Beyond Horndeski~\cite{Zumalacarregui:2013pma, Gleyzes:2014dya} or DHOST theories~\cite{Langlois:2015cwa}. The second is to remain within Horndeski theory while avoiding the no-go theorem by allowing the effective Planck mass to vanish in the asymptotic past. However, the latter approach raises concerns about unitarity violation, namely strong coupling at early times, which may undermine the theory's validity as an effective field theory.

Nevertheless, it is possible to identify regions of parameter space where the theory remains weakly coupled while still describing a non-singular Universe. We refer to such scenarios as non-pathological models with strong gravity in the past. Several examples of these models can be found in the literature; see Refs.~\cite{Ageeva:2021yik, Ageeva:2022asq, Akama:2022usl, GilChoi:2025hbs}. However, most of these constructions have difficulty producing a red-tilted scalar power spectrum, which is favored by current observational data~\cite{Planck:2018nkj,ACT:2025fju, ACT:2025tim}. In general, the unitarity bounds in these models are in tension with the red-tilted values required for the primordial scalar spectrum.

Several methods have been proposed to achieve a red-tilted spectrum. For example, Ref.~\cite{Akama:2022usl} fine-tunes the Lagrangian functions and sets certain terms in the cubic Lagrangian for perturbations to zero, thereby weakening the unitarity constraints and enabling a red-tilted spectrum. In Ref.~\cite{Ageeva:2022asq}, instead of relying on dimensional analysis, the authors derive exact unitarity bounds from the optical theorem. Using these exact bounds, they identify regions of parameter space where the theory remains weakly coupled and still yields a red-tilted scalar spectrum; this inevitably requires fine-tuning of model parameters. In Ref.~\cite{GilChoi:2025hbs}, a red-tilted scalar spectrum is obtained by introducing a spectator field.

These findings underscore the challenge of achieving a red-tilted scalar spectrum in non-singular cosmological models with strong gravity in the past. This motivates the interpretation of Genesis not as a replacement for inflation, but as a mechanism for setting the initial conditions of a conventional inflationary phase. In this framework, the Cosmic Microwave Background (CMB) modes freeze out during the inflationary epoch, while the Genesis phase serves as a pre-inflationary stage. Moreover, the existence of such a pre-inflationary Genesis era may introduce corrections to the inflationary dynamics, potentially resulting in observable signatures that distinguish this class of models from standard inflation.

In this paper, we construct a non-singular cosmological scenario consisting of two stages. The first stage is a Genesis phase, similar to that presented in Ref.~\cite{GilChoi:2025hbs}, and the second stage is classical Starobinsky inflation~\cite{Starobinsky:1980te, Martin:2013tda, Ketov:2025nkr}. Starobinsky inflation is equivalent to the Higgs inflation \cite{Bezrukov:2007ep} induced by so-called `non-minimal coupling' between the Ricci scalar and the inflaton~\cite{Futamase:1987ua, Park:2008hz,Hyun:2022uzc, Hyun:2023bkf} and its theoretical and phenomenological issues  have been extensively studied~\cite{Hamada:2014iga, Hamada:2014wna, He:2018mgb, Cheong:2022gfc, Cheong:2019vzl, Ema:2017rqn, Jinno:2019und, Koh:2023zgn, Cheong:2018udx,Lee:2020yaj,Park:2024ceu}. The model predicts the successful inflation with the spectral index $n_s=0.965$ and the tensor-to-scalar ratio $r=0.001$. (See \cite{Cheong:2021vdb} for a recent review.) However, these predictions met a challenge from the recent observation from ACT~\cite{ACT:2025fju, ACT:2025tim} in combination with Planck data, Lensing BAO data~\cite{Planck:2018jri} and the latest BICEP-KECK array B-model data~\cite{BICEP:2021xfz} :
\begin{align}
\left.n_s\right|_{\text{obs}}&= 0.975\pm 0.003, \,\,\,\,\,\,\left.r\right|_{\text{obs}}\lesssim 0.038. &\text{(P-ACT-LB-BK18)} 
\end{align}
Therefore, we would like to check the `correction' from our Genesis-Starobinsky setup to the vanilla Starobinsky model.

The Genesis phase of our scenario circumvents the Penrose singularity theorem \cite{Penrose:1964wq} due to an early violation of the null energy condition (NEC). Furthermore, the Borde–Guth–Vilenkin (BGV) theorem \cite{Borde:2001nh} does not apply, as our scenario belongs to specific subclasses identified in Ref. \cite{Lesnefsky:2022fen}, for which the BGV theorem is not applicable. We investigate whether a fully stable transition between these stages can be realized and analyze the corrections to Starobinsky inflation induced by the Genesis phase. Finally, we explore whether these Genesis-induced corrections can improve the consistency between Starobinsky inflation predictions and recent data from the Atacama Cosmology Telescope (ACT)~\cite{ACT:2025fju, ACT:2025tim}.

The rest of the paper is organized as follows. In Section~\ref{sec: The model}, we briefly describe the Starobinsky inflationary model and the early-time Genesis model. We construct the full Genesis-to-inflation scenario in Section~\ref{sec: Genesis to Inflation Scenario}. In addition, we perform numerical simulations and investigate the stability of the background solution. In Section~\ref{sec: Observational predictions}, we present the observational predictions of our model and compare them with the ACT data. We conclude in Section~\ref{sec: Conclusion}. In Appendix~\ref{App1}, we derive the connection between the lapse function value and the field value during the inflationary stage. In Appendix~\ref{App2}, we explicitly compute the leading-order Genesis correction to inflation. We present these corrections in two forms: the first as corrections to the potential, and the second as corrections in the form of $f(R)$ gravity.

\section{The model}
\label{sec: The model}

In this section, we briefly review both the Starobinsky inflationary model and the Genesis model.

The Lagrangian for the Starobinsky model is given by
\begin{align}
    \label{eq:Starobinsky_lagrangian}
    L_S = \frac{R}{2}    + \frac{  R^2}{12 M_{0}{}^2}\;,
\end{align}
where $M_0$ is a mass scale, and the Planck mass is set to unity, i.e., we work in Planck units ($8\pi G = 1/M_\mathrm{Pl}^2 = 1$).

The Lagrangian for the Genesis model from Ref.~\cite{GilChoi:2025hbs} at early times in the Jordan frame is written as:
\begin{align}
\label{genesisLagrangian}
\mathcal{L}_G &= G_2(\phi, X)-G_3(\phi, X)\square \phi
    + G_4(\phi)R\;,\\
    G_2 &= \frac{g X \left(- 3 c^2 e^{2 \phi } + 2 X\right) e^{\phi  (\delta +2 \mu -2)}}{3 c^4}+4 \mu ^2 X e^{2 \mu  \phi } \ln \left(\frac{X}{X_0}\right) \;, \nonumber\\
    G_3 &= \mu  e^{2 \mu  \phi } \left(\ln \left(\frac{X} {X_0}\right)+2\right) \;,\;\nonumber\\
    G_4 &= \frac{1}{2} e^{2 \mu  \phi } \;\nonumber,
\end{align}
where $X = -\frac{1}{2} \partial_\nu \phi \, \partial^\nu \phi$, and $\mu$, $\delta$, $c$ and $g$ are model parameters. The quantity $X_0$ is an arbitrary scale whose value is physically irrelevant, as rescaling $X_0$ only adds a total derivative to the Lagrangian.

In the ADM formalism (see Sec.~\ref{subsec: Model construction} for a brief review), the Lagrangian for the early Genesis phase has a much simpler form:
\begin{align*}
      L &= A_2(t, N) + A_3(t, N) K + A_4(t) (K^2 - K_{ij}^2) + B_4(t) R^{(3)}\;,\\
    A_2 &= \frac{1}{2} (-ct)^{-2\mu - 2 - \delta} \left( -\frac{g}{N^2} + \frac{g}{3 N^4} \right),\;
    A_3 = 0,\;
    A_4 = -\frac{1}{2} (-ct)^{-2\mu}\;.
\end{align*}

Typically, for the Genesis model in eq.~\eqref{genesisLagrangian}, the parameter $\mu$ is close to unity, while $\delta$ is close to zero. In the specific case where $\mu = 1$ and $\delta = 0$, the Lagrangian respects the global scale symmetry
\begin{align}
    \label{eq:scale_symm_for_phi}
    \tilde{\phi} &= \phi - \ln \lambda\;, \\
    \tilde{g}_{\rho\nu} &= \lambda^2 g_{\rho\nu}\;. \nonumber
\end{align}
It is interesting to note that the Starobinsky model \eqref{eq:Starobinsky_lagrangian} can be rewritten as
\begin{align}
    L_S = \frac{R}{2}  - \frac{e^{4\phi}}{12 M_0^2} + \frac{e^{2\phi} R}{6 M_0^2}\;.
\end{align}
Here, the first term breaks the symmetry, eq.~\eqref{eq:scale_symm_for_phi}, while the second and third terms preserve the same symmetry. Therefore, both the Starobinsky inflation and the Genesis scenario proposed in Ref.~\cite{GilChoi:2025hbs} are essentially built upon the same underlying principle: enforcing a global scale symmetry that is explicitly broken at a certain energy scale.

Finally, we would like to point out that the model proposed in Ref.~\cite{GilChoi:2025hbs} can be interpreted as a model with strong gravity in the asymptotic past, i.e., the effective Planck mass tends to zero as $t \to -\infty$. This raises the concern of a potential strong-coupling regime and unitarity violation in the early universe. However, as demonstrated in Ref.~\cite{GilChoi:2025hbs}, such issues do not arise in this particular Genesis scenario. For an appropriate parameter range [see eq.~\eqref{unitarity}], the strong-coupling scale always remains well above the energy scale of the background evolution. Moreover, the model remains consistent with the Weak Gravity Conjecture \cite{Arkani-Hamed:2006emk}. Specifically, we assume that at early times, normal matter is absent and the universe is dominated by a scalar field that drives the cosmological evolution. Importantly, the strong-coupling scale in the pure scalar sector is lower than that in the mixed and pure gravity sectors. As a result, scalar interaction terms dominate over both tensor–tensor and scalar–tensor interactions. Thus, gravitational interactions remain subdominant compared to scalar interactions---in other words, gravity remains the weakest force.

In the next section, we will demonstrate how to construct a fully stable cosmological scenario describing the Genesis phase \eqref{genesisLagrangian}, followed by Starobinsky inflation.

\section{Genesis to Inflation Scenario}
\label{sec: Genesis to Inflation Scenario}

\subsection{Model construction}
\label{subsec: Model construction}
 
As demonstrated in Ref.~\cite{GilChoi:2025hbs}, it is indeed possible to construct a Genesis scenario within the framework of generalized Galileon theories. This class of theories is equivalent to Horndeski gravity (see Ref.~\cite{Kobayashi:2011nu}), which represents the most general scalar–tensor theory with second-order equations of motion.

Following Ref.~\cite{GilChoi:2025hbs}, we will consider the following subclass of generalized Galileons:
\begin{align}\label{modelAction}
    \mathcal{S} = & \int d^4x \sqrt{-g} \left\{ G_2(\phi, X) - G_3(\phi, X) \square \phi + G_4(\phi) R \right\}, \quad X =  -\frac{1}{2} g^{\mu\nu} \partial_{\mu} \phi \partial_{\nu} \phi\;.
\end{align}
If we go to the Einstein frame, then this subclass is referred to as the Kinetic Gravity Braiding (KGB) model~\cite{Deffayet:2010qz}.

Moreover, similar to Refs.~\cite{Kobayashi:2016xpl, Ageeva:2021yik, GilChoi:2025hbs}, it is convenient both to construct the model and to perform matching between the Genesis and Starobinsky inflation in the ADM formalism.

Now, let us briefly review the ADM formalism. One writes the metric as follows:
\begin{equation}
\label{eq: FRW metric with Lapse}
    ds^2=-N^2 dt^2 +  
    \gamma_{ij}\left( dx^i+N^i dt\right)\left(dx^j+N^j dt\right)\;~, 
\end{equation}
where $N$, $N^i$ and $\gamma_{ij}$ are the lapse function, the shift vector, and the induced metric, respectively. Once we choose the unitary gauge, where $\phi$ depends only on $t$, the Lagrangian is written as ~\cite{Kobayashi:2019hrl}
\begin{align*}
    \mathcal{S} = \int d^4x \sqrt{-g} \left[ A_2(t, N) + A_3(t, N) K + A_4(t) (K^2 - K_{ij}^2) + B_4(t) R^{(3)} \right],
\end{align*}
where
\begin{equation*}
    A_4(t) = -B_4(t)~.
\end{equation*}
Here, we denote $\sqrt{-g} = N\sqrt{\gamma}$,
$K= \gamma^{ij}K_{ij}$, $^{(3)} R = \gamma^{ij} \phantom{0}^{(3)} R_{ij}$. Also, $K_{ij}$ is the extrinsic curvature of hypersurfaces $t=\mbox{const}$, which is given by
\begin{align*}
    K_{ij} &\equiv\frac{1}{2N}\left(\frac{d\gamma_{ij}}{dt}
    -\,^{(3)}\nabla_{i}N_{j}-\;^{(3)}\nabla_{j}N_{i}\right)~.
\end{align*}

The functions that appear in the covariant formalism and the ADM formalism are related by 
\begin{equation}
    G_2 = A_2 - 2X F_{\phi}, \quad G_3 = -2X F_X - F, \quad G_4 = B_4,
    \label{FromADMToCov}
\end{equation}
where the auxiliary function $F$ is given by
\begin{equation*}
    F_X = -\frac{A_3}{(2X)^{3/2}} - \frac{B_{4\phi}}{X}.
\end{equation*}
Here, $N$ and $X$ are related by
\begin{equation*}
    N^{-1} \frac{d\phi}{d t} = \sqrt{2X} ~.
\end{equation*}
The inverse relations are expressed by~\cite{DeFelice:2014bma} 
\begin{equation}\label{CovToADM}
\begin{split}
    A_2 &= G_2 - \sqrt{X}\int dX \frac{G_{3\phi}}{\sqrt{X}},\\
    A_3 &=- 2\sqrt{2 X} G_{4\phi}+\int dX G_{3X}\sqrt{2 X}, \\
    A_4 &= -G_4    
\end{split}
\end{equation}

Assuming a spatially flat FLRW universe, the equations of motion for the background spacetime are given by \cite{Kobayashi:2015gga}:
\begin{subequations}\label{backgroundEOM}
\begin{align}
    &(NA_2)_{N} + 3NA_{3N}H + 6N^2(N^{-1}A_4)_{N} H^2 = 0,\label{backgroundEOM1}\\
    &A_2 - 6A_4H^2 - \frac{1}{N} \frac{d}{d t}\left( A_3 + 4A_4H \right) = 0,\label{backgroundEOM2}
\end{align}
\end{subequations}
where $H(t) = N^{-1}\frac{d}{dt}\ln a(t)$ is the Hubble parameter.

Scalar perturbation $\zeta$ and tensor perturbation $h_{ij}$ about the flat FLRW background are described by the following quadratic actions~\cite{Kobayashi:2015gga}:
\begin{subequations}
\label{quadraticAction}
\begin{align}
    \label{quadraticActionScalar}
    \mathcal{S}_{\zeta \zeta}^{(2)} &= \int d t \, d^{3}x \, N a^3 \left[ \frac{\mathcal{G}_S}{N^2} \left( \frac{\partial \zeta}{\partial t} \right)^{2} - \frac{\mathcal{F}_S}{a^2} \left( \vec{\nabla} \zeta \right)^{2} \right] \;, \\
    \mathcal{S}_{hh}^{(2)} &= \int d t  \, d^3x \, \frac{N a^3}{8} \left[ \frac{\mathcal{G}_T}{N^2} \left( \frac{\partial h_{ij}}{\partial t} \right)^2 - \frac{\mathcal{F}_T}{a^2} h_{ij,k} h_{ij,k} \right] \;,
\end{align}
\end{subequations}
where
\begin{subequations}
\label{FS_GS}
\begin{eqnarray}
    \mathcal{F}_S &=& \frac{1}{a N} \frac{d}{d t} \left( \frac{a}{\Theta} \mathcal{G}_T^2 \right) - \mathcal{F}_T\;, \\
    \mathcal{G}_S &=& \frac{\Sigma}{\Theta^2} \mathcal{G}_T^2 + 3\mathcal{G}_T\;,
\end{eqnarray}
\end{subequations}
and
\begin{subequations}
\label{FT_GT}
\begin{align}
    \mathcal{G}_T &= -2A_4\;, \\
    \mathcal{F}_T &= 2B_4\;,
\end{align}
\end{subequations}
with
\begin{subequations}
\label{Sigma_Theta}
 \begin{align}
      \Sigma &= N A_{2N} + \frac{1}{2} N^2 A_{2NN} + \frac{3}{2} N^2 A_{3NN} H + 6H^2 A_4\;, \\
      \Theta &= 2H\left(\frac{NA_{3N}}{4H} - A_4\right)\;.
 \end{align}
\end{subequations}
Note that all the expressions above are obtained with the assumption $A_4 = A_4(t)$. For more general formulas, see Ref.~\cite{Kobayashi:2015gga}. The propagation speeds of scalar and tensor perturbations are given by
\begin{equation*}
   u_S^2 = \frac{\mathcal{F}_S}{\mathcal{G}_S}, \quad \text{and} \quad  u_T^2 = \frac{\mathcal{F}_T}{\mathcal{G}_T} = 1\;,
\end{equation*}
respectively.

We aim to construct a Lagrangian that reduces to Starobinsky inflation in the asymptotic future ($t\rightarrow \infty$) and approaches Genesis in the asymptotic past ($t\rightarrow -\infty$).

In the asymptotic past, we aim to realize the Genesis scenario, in which the Universe starts from flat spacetime and slowly expands. The Genesis solution, similar to that in  Ref.~\cite{GilChoi:2025hbs}, can be achieved by the following choice of Lagrangian functions $A_{2-4}$. First, we set $A^G_3=0$ in the asymptotic past. For the other functions, we choose a generic power-law behavior, similar to Refs.~\cite{Kobayashi:2016xpl,Ageeva:2021yik,GilChoi:2025hbs}:
\begin{equation}\label{ADMGenesis}
    \begin{split}
    A^G_2 &=  \frac{1}{2} (-c t)^{-2\mu - 2 - \delta} \cdot a_2(N) \\ 
    A^G_4  &= -\frac{1}{2} (-c t)^{-2\mu} \;~  ,
    \end{split}
\end{equation}
where $c$ is a positive constant with the dimension of mass.

One can show that the choice above in the asymptotic past allows the Genesis solution in the form of
\begin{subequations}\label{HubbleGenesis}
\begin{align}
    H(t) &\xrightarrow{t\rightarrow-\infty} h_0 \frac{1}{(- c t )^{1+\delta}}~,\\
    a(t) &\xrightarrow{t\rightarrow-\infty} a_g \left(1+\frac{h_0}{c \delta (- c t )^{\delta}}\right)~.
\end{align}
\end{subequations}
Here $a_g$ is an integration constant determined by the boundary condition in the asymptotic future. Substituting eqs.~(\ref{ADMGenesis}) and (\ref{HubbleGenesis}) into eq.~(\ref{backgroundEOM}) and extracting the leading-order contributions as $t\rightarrow -\infty$, one finds that
\begin{equation}
    a_2'(1) + a_2(1)=0
\end{equation}
from eq.~(\ref{backgroundEOM1}), and 
\begin{equation}
    h_0 \equiv   -\frac{a_2(1)}{4 c(1 +  \delta + 2 \mu)}  ,
\end{equation}
from eq.~(\ref{backgroundEOM2}), respectively. Here, we set $N(t=-\infty)=1$, utilizing time reparameterization invariance. In our Genesis model, $a_2(N)$ is specified as
\begin{equation}
    a_2(N) = -\frac{g}{N^2}+\frac{g}{3N^4}~,
\end{equation}
where the dimension of $g$ is $[g] = 4$.  
We see that at $|t| \sim c^{-1}$, the Genesis solution breaks down. Therefore, it is reasonable to assume that  the transition between Genesis and inflation occurs when $|t| \sim c^{-1}$.

Stable cosmological evolution during the Genesis regime is possible only within a limited range of the parameters $g,\;\mu$, and $\delta$. The allowed parameter space can be found by imposing a set of conditions. First of all, in order to circumvent the no-go theorem~\cite{Kobayashi:2016xpl, Libanov:2016kfc} and construct a fully stable Genesis scenario, one requires:
\begin{equation}\label{nogo}
    2\mu > 1 + \delta > 1~.
\end{equation}

Secondly, we require the Genesis regime to be described within classical field theory; in other words, one requires that the classical energy scale associated with the background solution is always well below the dynamical cutoff of the theory. The criterion to meet this requirement is given by  ~\cite{Ageeva:2018lko, Ageeva:2020gti}: 
\begin{equation}\label{unitarity}
    \mu + \frac{3}{2}\delta < 1. 
\end{equation}

Now, let us consider the asymptotic future. Using action \eqref{modelAction}, we find that the Starobinsky action in the Einstein frame~\cite{Coule:1987wt, Maeda:1987xf, Ketov:2025nkr} corresponds to 
\begin{equation}
    G^S_2 = X - V(\phi), \quad G^S_3 = 0, \quad G^S_4 = \frac{1}{2},
\end{equation}
where the scalar field potential is
\begin{equation}
    V(\phi) = \frac{3}{4}M_0^2\left[1 - \exp\left(-\sqrt{\frac{2}{3}}\phi\right)\right]^2~.
\end{equation}
We choose the following time slicing: 
\begin{align}
    \label{eq: time slicing}
    \phi(t) = \sqrt{\frac{2}{3}}[n - \ln (c t)],\;t>0\;.
\end{align}
After that, using eq. (\ref{CovToADM}), the functions $A_{2-4}$ for the Starobinsky model are given by \begin{equation}\label{ADMStarobinsky}
    \begin{split}
        A^S_2 &= -\frac{3 M_0^2}{4} + \frac{1}{3 N^2 t^2} + \frac{3 M_0^2 (c t)^{2/3}}{2 e^{2n/3}} - \frac{3(ct)^{4/3} M_0^2}{4 e^{4n/3}},\\
        A^S_3 & = 0,\\
        A^S_4 & = -\frac{1}{2}.
    \end{split}
\end{equation}
Here, $c$ is a positive constant with dimensions of mass.
We introduced a new parameter $n$ that determines the characteristic value of $\phi$ at the end of the Genesis-to-inflation transition. Let us clarify the meaning of the parameter $n$. The coordinate time $t$ can be expressed as
\begin{align*}
    t = \frac{1}{c} \exp\Big[\sqrt{\tfrac{3}{2}}\Big( \sqrt{\tfrac{2}{3}} n - \phi\Big)\Big]\;.
\end{align*}
The transition occurs approximately at $t \sim c^{-1}$, corresponding to a field value $\phi \sim \sqrt{2/3} \, n$.  
Hence, $n$ sets the typical field value during the transition.  
Larger $n$ implies a transition further from the epochs relevant to the Cosmic Microwave Background (CMB), so for sufficiently large $n$, the Genesis corrections become negligible.  
This expectation will be confirmed by the numerical simulations in Section~\ref{sec: Observational predictions}.

In order to match the early-time Genesis stage and Starobinsky inflation, we introduce two auxiliary functions:  
\begin{equation}
    f(t) = \frac{c}{2}\left(\frac{\ln [2\cosh(s t)]}{s}-t\right)+1
\end{equation}
and
\begin{equation}
    U(t) = \frac{e^{s t}}{e^{s t}+1}~.
\end{equation}
The parameter $s$ controls the time scale of the transition phase. Note that their asymptotic behaviors are
\begin{equation*}
    f(t)\xrightarrow{t\rightarrow -\infty} - c t~,\quad f(t)\xrightarrow{t\rightarrow \infty} 1    
\end{equation*}
and
\begin{equation*}
    U(t)\xrightarrow{t\rightarrow -\infty} 0~,\quad U(t)\xrightarrow{t\rightarrow \infty} 1~.
\end{equation*}
As we will see below, the function $f$ plays the role of the effective Planck mass, while the function $U$ is responsible for the suppression of higher-order derivative terms and governs the transition between Genesis and inflation.

Using these functions, the full Lagrangian can be written as
\begin{equation}\label{fullLagrangian1}
\begin{split}
    A_2 = \frac{1}{2} &f(t)^{-2\mu-2-\delta}\left(-\frac{g}{N^2}+\frac{g}{3N^4}\right)(1-U(t)) \\
    &+ \left(-\frac{3 M_0^2}{4} + \frac{1}{3 N^2 \tau(t)^2} + \frac{3 M_0^2 (c \tau(t))^{2/3}}{2 e^{2n/3}} - \frac{3(c\tau(t))^{4/3} M_0^2}{4 e^{4n/3}}
\right)U(t),
\end{split}
\end{equation}
where $\tau(t) = 2f(t)/c+t$, and
\begin{equation}\label{fullLagrangian2}
    A_4 = -\frac{1}{2} f(t)^{-2\mu}~.
\end{equation}
One can see that these functions satisfy
\begin{equation*}
    A_{2,4} \xrightarrow{t\rightarrow -\infty} A^G_{2,4},\quad \text{and} \quad A_{2,4} \xrightarrow{t\rightarrow \infty} A^S_{2,4}.
\end{equation*}
Therefore, the setup smoothly matches the Genesis scenario in the asymptotic past and Starobinsky inflation in the asymptotic future.

\subsection{Numerical simulation}
\label{subsec: Numerical simulation}

Here, we explore the evolution of our model by solving the equations of motion numerically. We show that it allows a stable transition from the Genesis phase to the Starobinsky inflation phase.

To demonstrate this, let us choose the following  model parameters:
\begin{equation}\label{parameters}
\begin{split}
    \mu = 0.7~,\quad \delta = 0.1~,\quad &c=5\times10^{-5},\quad s=2\times10^{-5}~,\quad g=5\times 10^{-8} ~,\\
    &\quad M_0 = 10^{-5} ~,\quad n=10.4~.    
\end{split}
\end{equation}

Let us briefly comment on the rationale behind our choice of parameters. The values of $\mu$ and $\delta$ are chosen from the parameter space constrained by eqs.~(\ref{nogo}) and (\ref{unitarity}), in order to evade both the no-go theorem and the strong-coupling problem. The mass parameter $M_0$ is chosen to match the amplitude of the scalar power spectrum as measured by CMB observations~\cite{Starobinsky:1983zz, Planck:2018nkj}. Consequently, the late-time expansion rate becomes $H \sim M_0/2$. We set the parameters $c$ and $s$ close to the $M_0$ scale, i.e., $c \sim s \sim M_0$.

Next, we choose the parameter $g$ to be roughly of the same order as $M_{\text{Pl}}^2 c^2$. Specifically, for our current choice, we have $\frac{g}{c^2 M_{\text{Pl}}^2} \sim 10^1$ (here, for clarity, we temporarily restore the Planck mass). In other words, we adopt a natural parameter choice in which all quantities are approximately of the same order of magnitude. Finally, with this set of parameters, the time scale of the transition stage remains much shorter than those of the Genesis and inflationary stages.

The equation of motion, eq.(\ref{backgroundEOM}), can be solved by dividing the evolution into four distinct regimes: (i) early power-law Genesis, (ii) non-power-law Genesis, (iii) transition phase, and (iv) Starobinsky inflation. This approach is motivated by Section 5 of Ref.\cite{GilChoi:2025hbs}.

\begin{itemize}
    \item \textbf{Early power-law Genesis}
    
     For the early phase $-\infty < t \ll -c^{-1}$, we assume a power-law ansatz 
    $$H(t)=h_0(-c t)^{-1-\delta} \left( 1 +  O\big[(-ct)^{-\delta}\big]\right).$$ 
    Inserting this ansatz into eq. (\ref{backgroundEOM}), eq. (\ref{backgroundEOM1}) becomes
    \begin{equation}
    \frac{1}{2}g (- c t)^{-2-\delta -2\mu}(-N^{-4}+N^{-2}) + 3 h_0^2(- c t)^{-2-2\delta -2\mu}=0~,
    \end{equation}
    and eq. (\ref{backgroundEOM2}) becomes
    \begin{equation}
     \frac{1}{6}(- c t)^{-2-\delta -2\mu}(g N^{-4}-3gN^{-2}+12 c h_0 (1+\delta+2\mu)N^{-1}) + 3 h_0^2(- c t)^{-2-2\delta -2\mu}=0 ~.    
    \end{equation}
    Note that at early times we assume $f(t)\sim -c t$ and $U(t)\sim 0$; hence, $A_2\sim A^G_2$ and $A_4\sim A^G_4$ when $t \to -\infty$.
    One can see that the solution has the following form
    \begin{equation}\label{eq: sol Early Genesis HN}
        N=1 + O\big[(-ct)^{-\delta}\big]~, \quad h_0=\frac{g}{6c(1+\delta+2\mu)}.
    \end{equation}
    This solution is used to determine the initial condition for the following non-power-law Genesis regime.

    \item \textbf{Non-power-law Genesis}
    
    As the evolution approaches the transition phase around $-t \sim c^{-1}$, the power-law approximation becomes increasingly inaccurate due to neglected $O[(-ct)^{-\delta}]$ corrections. The time at which the power-law description breaks down was estimated in Ref.~\cite{GilChoi:2025hbs}, and this time $t_{nl}$ is given by:
    \begin{align*}
        t_{nl} \sim c^{-1} \left( \frac{g}{c^2 M_\mathrm{Pl}^2} \right)^{1/\delta}
    \end{align*}
 Here, for clarity, we temporarily restore the Planck mass.
 Therefore, for $-t < t_{nl}$, a more accurate description than the power-law one is required. Note that $t_{nl}\sim 10^{13}c^{-1}$ with our parameter choice, which is given by eq.~(\ref{parameters}), at this time $-t \sim t_{nl}$, it is still legitimate to use the following asymptotics  $f(t)\sim -c t$ and $U(t)\sim 0$ . In this case, by introducing new variables
    \begin{equation*}
        u = (- c t)^{-\delta}\quad\text{and}\quad h = u^{-1-1/\delta}H N~,
    \end{equation*}
    one can rewrite eq.~(\ref{backgroundEOM1}) as
    \begin{equation}
        \frac{1}{2}u^{(2+\delta+2\mu)/\delta} (-gN^{-4}+(g+6u h^2)N^{-2})=0~.
    \end{equation}
    Its solution can be found as
    \begin{equation}\label{Nusol}
        N= \frac{1}{\sqrt{1+6 u h^2/g}}~.
    \end{equation}
    One can also rewrite eq.~(\ref{backgroundEOM2}) as  
    \begin{equation}
    \begin{split}
        \frac{1}{6}u^{(2+\delta+2\mu)/\delta} \bigg[&gN^{-4}-12c\delta uh\frac{dN}{du}N^{-3}\\
        &+3\left\{-g+4c(1+\delta+2\mu )h+6uh^2+4c\delta u \frac{dh}{du}\right\}N^{-2}\bigg]=0~,
        \end{split}
    \end{equation}
     where $N = N(u)$ and $h = h(u)$ are now functions of variable $u$ and $N(u)\big|_{u \to 0} \to 1$.
     Using eq.~(\ref{Nusol}), we solve the above equation for $h(u)$ numerically from an initial value  $u=10^{-8}$, satisfying $u\ll(c t_{nl})^{-\delta}=0.05$. The initial condition for $h(u)$ is taken from the power-law solution.

     \item \textbf{Transition phase}

    In the time range $|t| < c^{-1}$, the transition from the Genesis phase to the Starobinsky inflation phase occurs. During the transition stage, $A_2$ and $A_4$ is given by eqs.~(\ref{fullLagrangian1})--(\ref{fullLagrangian2}), and the behavior of these functions is quite non-trivial. Therefore, we rely on a numerical method to obtain an accurate solution for the Hubble parameter $H(t)$. Note that eq.~(\ref{backgroundEOM1}) is purely algebraic and can be solved analytically. This equation gives the connection between the Hubble parameter and the lapse function at any given moment in time. We select the largest real positive root among the four roots of eq.~(\ref{backgroundEOM1}). 
    This is the only root which leads to the self-consistent evolution of the system. Namely, the lapse function, Hubble parameter, and their first derivatives are continuous functions during the whole evolution. 
    Afterwards, we substitute the analytical solution for the lapse function into the second equation (\ref{backgroundEOM2}) and numerically evolve $H(t)$.

     We start the transition phase simulation from $t=-100c^{-1}$. At this time, the effects of terms responsible for the transition are still negligible, thus, we can safely make the matching between stages and set the initial conditions.  The $H$ value at this point is taken from the numerical solution of the previous regime. We continue the evolution of $H(t)$ until the system almost behaves like Starobinsky inflation. 

     \item \textbf{End of transition stage and Starobinsky inflation}

    As the phase of evolution enters the Starobinsky inflationary regime, solving the equation of motion using the ADM coordinate time becomes inefficient when the lapse function decays. It can be shown that the lapse function is given by
    \begin{equation}
    \label{eq: N from phi}
        N \simeq \frac{c}{M_0}e^{- n + \frac{5}{\sqrt{6}}, \phi}\;,
    \end{equation}
    where we assume the time-slicing is given by eq.~\eqref{eq: time slicing}. We also assume that the evolution is described by the Starobinsky inflation model. See Appendix~\ref{App1} for a derivation of eq.~\eqref{eq: N from phi}.
         
    Eq.~\eqref{eq: N from phi} shows that the lapse function rapidly decays as the field $\phi$ rolls down the potential $V(\phi)$, and as a result, the evolution of cosmic time 
    \begin{equation}
        dt_c =N dt
    \end{equation}
    effectively freezes. Therefore, in our simulation, we evolve the system using the ADM time coordinate until $t = 800c^{-1}$, and then continue the evolution using the covariant formalism, namely, we use the following equation
    \begin{equation}\label{StarobinskyEOM}
         \frac{d^2\phi}{d t_c^2} + 3 H\frac{d\phi}{d t_c} + V'(\phi)=0~.
    \end{equation}    
     
     The initial conditions for eq.~(\ref{StarobinskyEOM}) can be obtained from the previous regime by using the gauge choice for the $\phi(t)$ field and the expression for the first derivative of the scalar field
     \begin{equation*}
         \frac{d\phi}{dt_c} =-\sqrt{\frac{2}{3}} \frac{1}{tN}~.
     \end{equation*}
     We terminate the evolution when the (Hubble) slow-roll parameter $\epsilon_H\equiv -\dot{H}/H^2$ roughly reaches unity.
\end{itemize}

\begin{figure}
    \centering
    \includegraphics[width=0.47\linewidth]{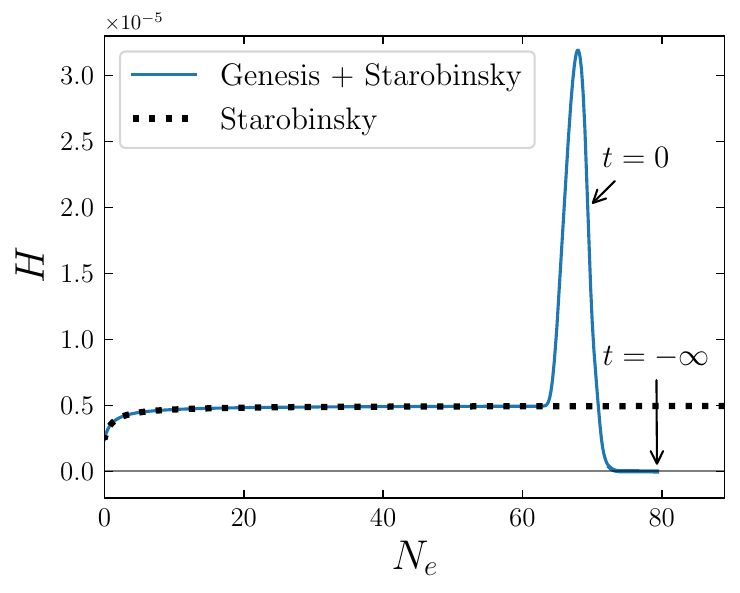}
    \includegraphics[width=0.45\linewidth]{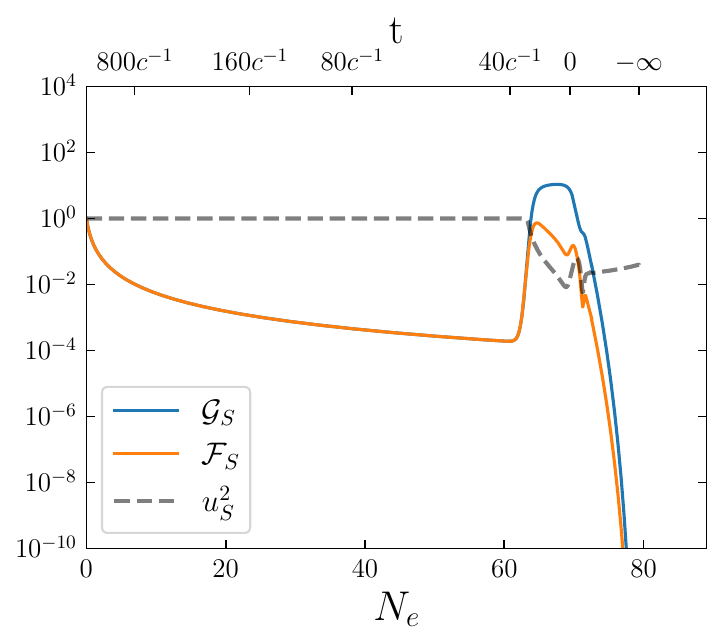}
    \caption{ (Left) Evolution of $H$ in our model ($\text{Genesis to inflation}$) and in the Starobinsky model as a function of $N_e$. The values of $H$ at the beginning of the universe ($t=-\infty$), and at the moment of transition ($t=0$) are indicated with arrows. (Right) Evolution of $\mathcal{G}_S$, $\mathcal{F}_S$, and $u_S^2$ as a function of $N_e$. ADM time coordinates at selected points are shown along the top horizontal axis. In both panels, the evolution is computed with the parameter values
    $\mu = 0.7$, $\delta = 0.1$, $c=5\times10^{-5}$, $s=2\times10^{-5}$, $g=5\times 10^{-8}$, $M_0 = 10^{-5}$, and $n=10.4$.}
    \label{fig:evolution}
\end{figure}

The evolution results of $H$, $\mathcal{G_S}$, $\mathcal{F}_S$ and $u_S^2$ are presented in figure~\ref{fig:evolution}. Here, we parameterize them with the e-fold number \textit{before} the end of inflation:
\begin{equation}
    N_e(t) = \int_{a}^{a(t_\mathrm{end})}d\ln a= \int_t^{t_{end}} H(t) N(t)dt~,
\end{equation}
where $t_\mathrm{end}$ is the moment of time at which $\epsilon_H(t_\mathrm{end}) \sim 1$. Moreover, one can estimate the total number of e-folds, $N_\text{max}$, during the entire evolution—from the beginning of the Universe to the end of inflation—as follows. First, during the early Genesis phase, one uses the relation between the variable $u$ and time, $u = (-ct)^{-\delta}$, and obtains: 
\begin{equation*}
    \int H(t) N(t) dt = (c \delta)^{-1}\int  h(u) du~.
\end{equation*}

In the early power-law Genesis phase, the integral further reduces to  
\begin{equation*}
    (c \delta)^{-1}\int  h(u) du \simeq (c \delta)^{-1} h_0 u ~.
\end{equation*}
Finally, we arrive at the expression for $N_\mathrm{max}$
\begin{equation}\label{Ne}
    N_\mathrm{max} =    \left[\int_{t_i}^{t_{end}} H(t)N(t) dt+(c\delta)^{-1}\left(\int_{u_0}^{u_i} h(u)du + h_0 u_0\right)\right]~,
\end{equation}
where $u_0=10^{-8}$, $u_i=(-ct_i)^{-\delta}$, $t_i=-100c^{-1}$ . Also, at minus infinity $N_\mathrm{max}=N_e(-\infty)$. Thus, $N_\mathrm{max}$ is the maximum value of $N_e$, or equivalently, $N_e$ at the beginning of the universe. For our model, we roughly have $N_\mathrm{max}\simeq80$ with the parameter choice eq.~(\ref{parameters}).

The left panel of figure~\ref{fig:evolution} presents the evolution of the Hubble parameter, starting from the Genesis phase, passing through the transition around $t \simeq 0$, and entering the Starobinsky inflation regime. For comparison, the Hubble parameter evolution of Starobinsky inflation is also plotted in the same panel. The curve of the full Genesis scenario is seen to converge to that of the Starobinsky inflation case for $N_e \lesssim 60$. Although the two curves are nearly aligned after the transition phase, their small deviation can still give rise to differences in observables, which will be discussed in the next section.

The right panel of figure~\ref{fig:evolution} shows our model remains stable, i.e., $\mathcal{G}_S, \mathcal{F}_S > 0$, from the beginning of the universe to the end of inflation. The same panel also presents the sound speed $u_S$, which remains sub-luminal during the Genesis and approaches $u_S=1$ after the transition phase ($t>0$). Stability at the early power-law Genesis phase is ensured by our parameter choice eq.~\eqref{parameters}. Stability of the intermediate phases---the non-power-law Genesis and the transition phase---is highly nontrivial and depends on the shape of transition functions $f$ and $U$. Nevertheless, our numerical results for $\mathcal{G}_S$ and $\mathcal{F}_S$ show that stability can still be maintained. The evolution during inflation is also stable, as expected from the expressions:
\begin{equation}
    \mathcal{G}_S = \mathcal{F}_S \simeq -\frac{1}{H^2} \frac{dH}{dt_c} > 0~,\quad u_S = 1 \quad (\text{during inflation})~.
\end{equation}
This formula is the standard one and, for example, can be found in Ref.~\cite{Maldacena:2002vr}.

\section{Observational predictions}
\label{sec: Observational predictions}

\begin{figure}
    \centering
    \includegraphics[width=0.5\linewidth]{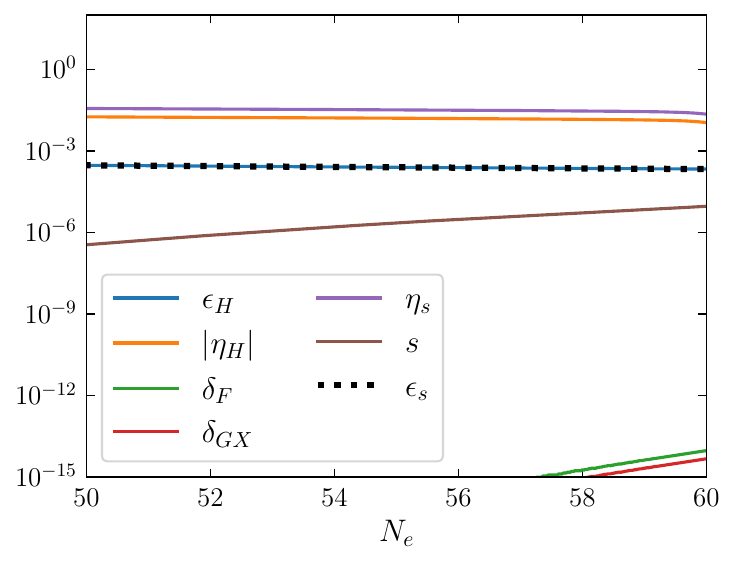}
    \caption{ Example of the slow-roll and the slow-variation parameters for $50<N_e<60$. We use the same parameter values as in figure~\ref{fig:evolution}. 
    }
    \label{fig:slowroll}
\end{figure}

Our model, in which Genesis serves as the initial condition for Starobinsky inflation, leads to corrections to the cosmological observables predicted by the standard inflationary model. In particular, it tends to predict larger values of the scalar spectral index $n_s$ and the tensor-to-scalar ratio $r$ compared to the standard scenario. Notably, recent ACT data~\cite{ACT:2025fju, ACT:2025tim} disfavor the vanilla Starobinsky model at $\gtrsim 2\sigma$, as they measure a slightly larger $n_s$ than predicted for 50--60 e-folds of inflation. This mismatch has prompted considerable discussion about what modifications to Starobinsky inflation might be necessary~\cite{Drees:2025ngb, Addazi:2025qra, Dioguardi:2025mpp, He:2025bli}. In contrast, our model is motivated by a fundamental question: is it possible to construct a non-singular cosmological scenario whose late-time behavior approaches that of Starobinsky inflation? Interestingly, Genesis-Starobinsky setup predicts a shift in the scalar spectral index toward larger values, improving consistency with ACT results. Thus, our findings offer a fresh and insightful perspective on how this observational tension might be interpreted.

We obtain the observables, $n_s$ and $r$, using the formulas given in Ref.~\cite{DeFelice:2011zh}. These formulas represent a generalization of the slow-roll approach for modified gravity models. 
The values of $n_s$ and $r$ are expressed in terms of the \textit{slow-variation} parameters:
\begin{equation}\label{nsandrAntonio}
    n_s = 1 - 2 \epsilon_H - \delta_F - \eta_s - s ~,  \quad
    r = 16 u_S \epsilon_s~,
\end{equation}
where the slow-variation (roll) parameters are
\begin{equation}\label{slowv1}
\begin{split}    
    \epsilon_s &= \epsilon_H + \frac{1}{2}\delta_F + \delta_{GX}\\
    \eta_s & = \frac{\dot{\epsilon_s}}{H\epsilon_s}\\
    s &= \frac{\dot{u_S}}{Hu_S}    
\end{split}
\end{equation}
and 
\begin{equation}\label{slowv2}
    \delta_F = \frac{\dot{G_4}}{HG_4 }~, \quad \delta_{GX} = \frac{\dot{\phi }X G_{3X}}{2 H G_4} ~.
\end{equation}
Here, an overdot denotes the derivative with respect to cosmic time, $d/dt_c$. Since $B_4$ in our model is a function of time only, the following relation between the covariant functions holds:
 $$G_{3X} = G_{4\phi}/X,$$ see eq.~(\ref{FromADMToCov}). 
Also, one can obtain
\begin{equation*}
    \delta_{GX} = \frac{\dot{\phi}G_{4\phi}}{2HG_4} = \frac{\dot{G_4}}{2HG_4} = \frac{1}{2} \delta_F~.
\end{equation*}

It is interesting to note that sufficiently long after the transition phase, our model essentially reduces to slow-roll inflation on the potential, i.e., $G_2 = X - V(\phi)$, $G_3 = 0$, and $G_4 = 1/2$. Therefore, the slow-roll parameters can be roughly estimated as:  
\begin{align*}
    \epsilon_s &\simeq \epsilon_H \text{ and } \eta_s \simeq -2\eta_H + 2\epsilon_H,\\
    \epsilon_H &= -\frac{\dot{H}}{H^2} = \frac{\dot{\phi}^2}{2 H^2} \simeq \frac{1}{2}\left(\frac{V^{\prime}(\phi)}{V(\phi)}\right)^2~,\\
    \eta_H &= -\frac{\ddot{H}}{2\dot{H}H} = -\frac{\ddot{\phi}}{H\dot{\phi}} \simeq \frac{V^{\prime\prime}(\phi)}{V(\phi)}~.
\end{align*}
Accordingly, eqs.~\eqref{nsandrAntonio} and \eqref{slowv2} are in full agreement with standard inflationary calculations.

Here we focus on the case where the modes relevant to CMB observations freeze out at $50 < N_e < 60$. One can estimate $N_e$ at the time a $k$-mode is frozen as follows:
\begin{equation}
\begin{split}
    N_e &= \ln \frac{a_\mathrm{end}}{a_f}\\ 
    &= \ln \frac{a_\mathrm{end}}{a_\mathrm{reh}}\frac{a_\mathrm{reh}}{a_{0}}\frac{a_0 H_f}{a_f H_f} \\
    &= -N_\mathrm{reh} - N_\mathrm{BB} + \ln \frac{H_f}{k_*}    
\end{split}
\end{equation}
where we set $k_* = a_f H_f$ and $a_0 = 1$. Here, $N_\mathrm{reh} = \ln (a_\mathrm{reh}/a_\mathrm{end})$ is the number of e-folds during the reheating phase, and $N_\mathrm{BB} = \ln (a_0 / a_\mathrm{reh})$ is the number of e-folds from the end of reheating to the present universe. 
We take the following parameter values: 
\begin{equation}
\begin{split}
    k_* &= 0.05 ~\mathrm{Mpc^{-1}} = 2.6\times10^{-59}~\\    
    H_f & \simeq M_0/2 = 5.0\times10^{-6}
\end{split}    
\end{equation}
where all values are given in Planck units. Then, one arrives at
\begin{equation}
    N_e \simeq 123 - N_\mathrm{reh} - N_\mathrm{BB}~.
\end{equation}

To estimate $N_\mathrm{BB}$, we use 
\begin{equation}
    N_{BB} \simeq \ln \frac{T_\mathrm{reh}}{T_0}~,
\end{equation}
where 
\begin{equation}
    T_0=2.725 ~\mathrm{K}=10^{-32}~M_\mathrm{Pl}~,
\end{equation}
and $T_\mathrm{reh}$ is the temperature during the reheating. Here, $M_\mathrm{Pl}=1/\sqrt{8\pi G}\simeq 2.4\times10^{18}~\mathrm{GeV}$ is Planck mass. If we assume a nearly instantaneous reheating, i.e, $N_\mathrm{reh} \simeq 0$, and take the reheating temperature $T_\mathrm{reh}\simeq 10^{-6}~M_\mathrm{Pl}\sim 10^{13}~\mathrm{GeV}$, then we have $N_\mathrm{reh}+N_{BB}\simeq 60$, which leads to $N_e\simeq 63$. However, the resulting $N_e$ can be larger or smaller depending on $T_\mathrm{reh}$ and the detailed dynamics of the reheating process. Therefore, we assume that our model allows mode freezing within the commonly considered range, $50<N_e<60$. Moreover, we will assume that the 
mode freeze occurs sufficiently far from the transition point meaning that Genesis corrections to Starobinsky inflation are small. There exist other interesting possibilities---such as the mode exiting the horizon during the transition ($|t| \lesssim c^{-1}$); however, this possibility is beyond the scope of this work and is left for future studies.

Using eq.~(\ref{nsandrAntonio}) is justified only when the slow-variation parameters remain small. We find that, as long as the Genesis-to-Starobinsky transition ($t \sim 0$) occurs at $N_e > 60$, this condition is satisfied in most cases. The behavior of the slow-roll parameters for the parameter choice given in eq.~(\ref{parameters}) is shown in figure~\ref{fig:slowroll}. In this example, the transition occurs around $N_e = 70$ (see figure~\ref{fig:evolution}), and, as a result, both the slow-variation and slow-roll parameters remain small within $50 < N_e < 60$.

Our model introduces several new parameters: $\mu$, $\delta$, $c$, $s$, $g$, and $n$. In this work, we focus solely on $g$ and $n$ for simplicity. Accordingly, throughout the remaining sections, we fix the following parameters unless otherwise specified:
\begin{equation}\label{parfixed}
    \mu = 0.7~,\quad \delta = 0.1~,\quad c=5\times10^{-5},\quad s=2\times10^{-5}~,\quad M_0=10^{-5}~.
\end{equation}

\begin{figure}
    \centering
    \includegraphics[width=0.45\linewidth]{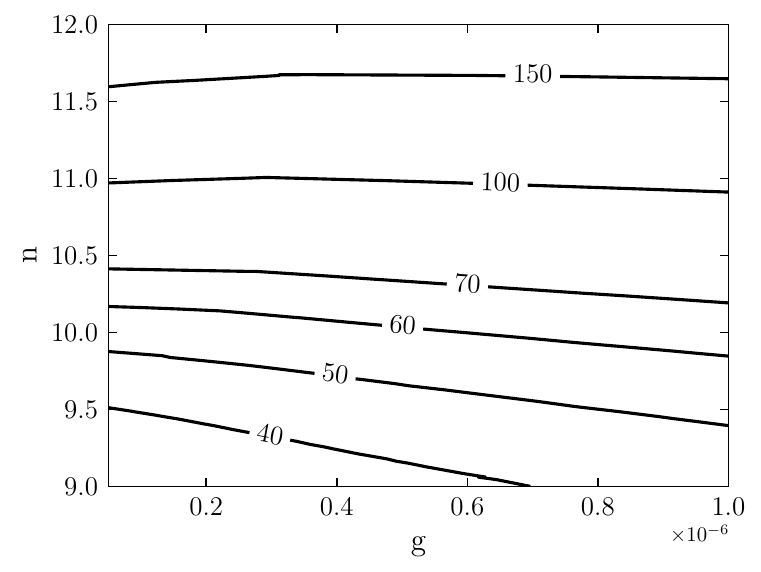}
    \caption{Contours of $N_\mathrm{max}$ in the $g$-$n$ parameter space. The other model parameters are fixed as $\mu = 0.7$, $\delta = 0.1$, $c=5\times10^{-5}$, $s=2\times10^{-5}$, and $M_0=10^{-5}$.}
    \label{fig:Nmax}
\end{figure}

\begin{figure}
    \centering
    \includegraphics[width=0.45\linewidth]{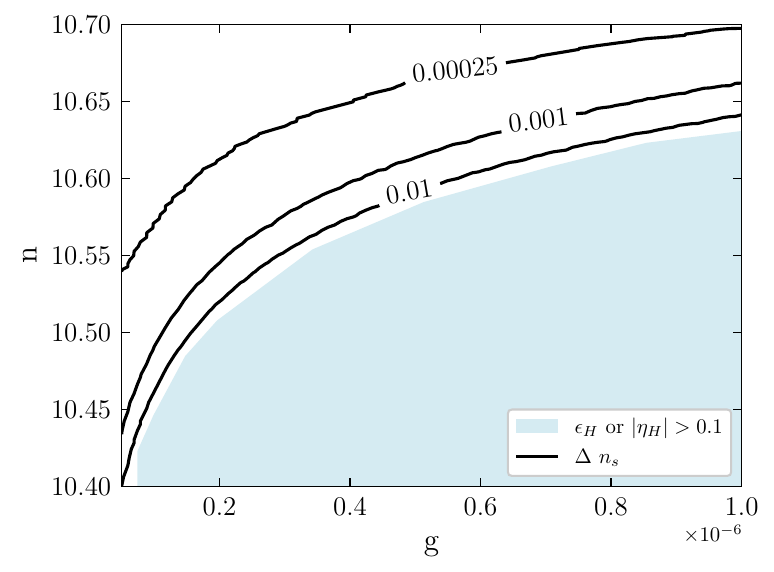}
    \caption{ Contours of $\Delta n_s=n_s-n_s^\text{st}$ which is evaluated at $N_e=60$. $n_s$ is from our model, and $n_s^{st}$ is from the Starobinsky model. The blue region is excluded due to violation of the slow-roll condition $\epsilon_H,|\eta_H|<0.1$. The same parameter values as in figure~\ref{fig:Nmax} are used. }
    \label{fig:delta}
\end{figure}

Now let us briefly comment on how the physical quantities depend on the model parameters $(n,g)$. First of all, the maximum e-fold number, $N_\mathrm{max}$, rapidly increases with increasing $n$, as shown in figure~\ref{fig:Nmax}. This is because larger $n$ corresponds to a larger initial value of $\phi$, which results in a longer inflationary phase. Although larger $g$ also induces a larger $N_\mathrm{max}$, its effect turns out to be subdominant. This can be understood from $\Delta N_e \propto h_0 \propto g$ (see eq.~(\ref{Ne})), which shows that $g$ primarily controls the e-fold number of the Genesis phase. Overall, figure~\ref{fig:Nmax} shows that $n \lesssim 9.5$ becomes irrelevant to our considerations, since $N_\mathrm{max} < 50$.

We compute $\Delta n_s$, defined as the difference between $n_s$ predicted by our model and that from the Starobinsky model. The results are shown as contour plots in the $g$-$n$ parameter space in figure~\ref{fig:delta}. Similar to the case of $N_\mathrm{max}$, $n$ has a greater impact on $\Delta n_s$. Additionally, we find that $\Delta n_s$ \textit{always} takes positive values, which is in \textit{agreement} with recent ACT data~\cite{ACT:2025fju, ACT:2025tim}.

It is interesting to note from figure~\ref{fig:delta} that $n_s$ begins to change very rapidly as one approaches the blue region. This blue region corresponds to the part of parameter space for which the slow-roll conditions,
\begin{equation}\label{slowroll}
    \epsilon_H < 0.1~, \quad \text{and} \quad |\eta_H| < 0.1~,
\end{equation}
are violated. This behavior arises from the rapid change of the Hubble parameter during the Genesis-to-inflation transition. In other words, the deviation from the Starobinsky solution becomes significantly large.

\begin{figure}
    \centering
    \includegraphics[width=0.5\linewidth]{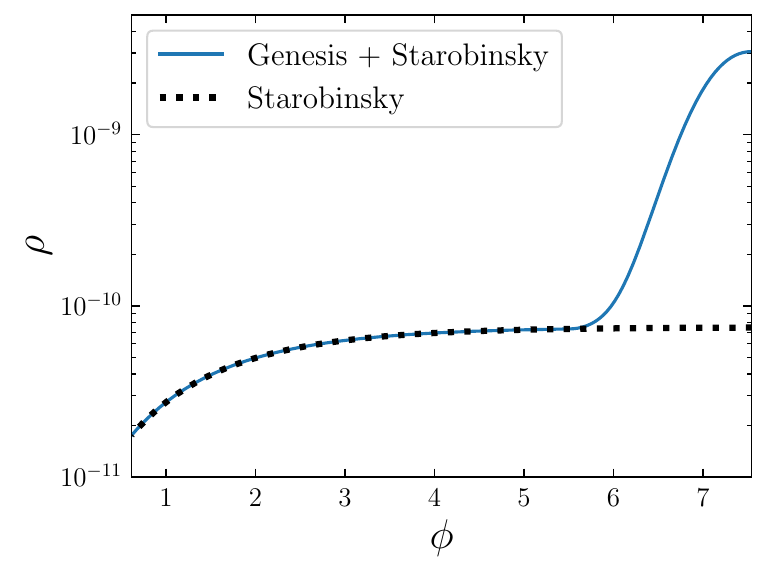}
    \caption{Illustrative example of $\rho$ as a function of $\phi$, comparing our model (solid curve) with the Starobinsky model (dotted curve). The same parameter values as in figure~\ref{fig:evolution} are used.
    }
    \label{fig:rho}
\end{figure}

In order to demonstrate this explicitly, it is instructive to show the behavior of the effective energy density, $\rho(\phi) = 3H^2$. Of course, for non-minimal coupling, the definition of energy density involves some ambiguity, since gravity and the scalar field cannot be completely separated. Thus, following Ref.~\cite{GilChoi:2025hbs}, we define the effective energy density as the 00-component of $R^\mu_\nu - \frac{1}{2}\delta^\mu_\nu R$. We use this definition to compute the energy density shown in figure~\ref{fig:rho}. We see that, as one approaches the transition point, the deviation in energy density between the Starobinsky model and the Genesis-to-inflation model grows significantly.

Moreover, if one assumes that corrections from Genesis to inflation are sufficiently small, the leading-order correction reduces to a modification of the inflationary potential, while all other higher-order terms (involving higher derivatives) are suppressed. In this case, inflation still occurs within the slow-roll regime; however, the potential receives the following corrections: 
\begin{align}
    \label{eq: psi model}
    L_{\psi} &= X_{\psi} - V_{\psi}(\psi)\,,\\
    V_{\psi}(\psi) &= \frac{3}{4} M_0^2 e^{-2 \sqrt{\frac{2}{3}} \psi }
    \left(e^{\sqrt{\frac{2}{3}} \psi }-1\right)^2 + \frac{3 g M_0^2 \left(e^{\sqrt{\frac{2}{3}} \psi }-1\right)
    e^{-\frac{s e^{n-\sqrt{\frac{3}{2}} \psi }}{c}-\frac{7 \psi }{\sqrt{6}}} \left(c e^{\sqrt{\frac{3}{2}} \psi }+e^n s\right)}{4 c s^2} \nonumber
\end{align}
For more details, see Appendix~\ref{App2}. It is noteworthy that the leading-order correction does not take the form of polynomial terms in powers of $R$ (see Appendix~\ref{App2}). Therefore, our model is \textit{novel} and fundamentally different from the Starobinsky model with $\sum_i c_i R^i$ corrections, which has been investigated, for example, in Refs. ~\cite{Cheong:2020rao, Rodrigues-da-Silva:2022qiq, Kim:2025dyi}.

\begin{figure}
    \centering
    \includegraphics[width=0.45\linewidth]{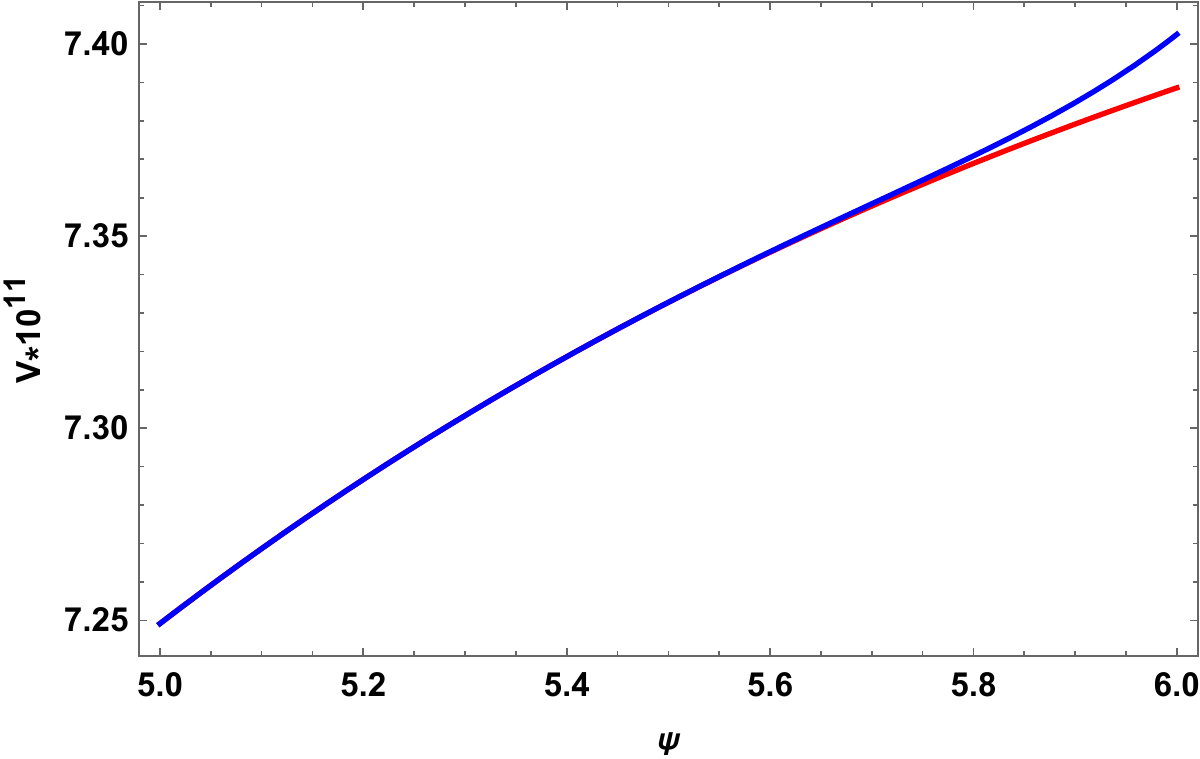}
    \includegraphics[width=0.45\linewidth]{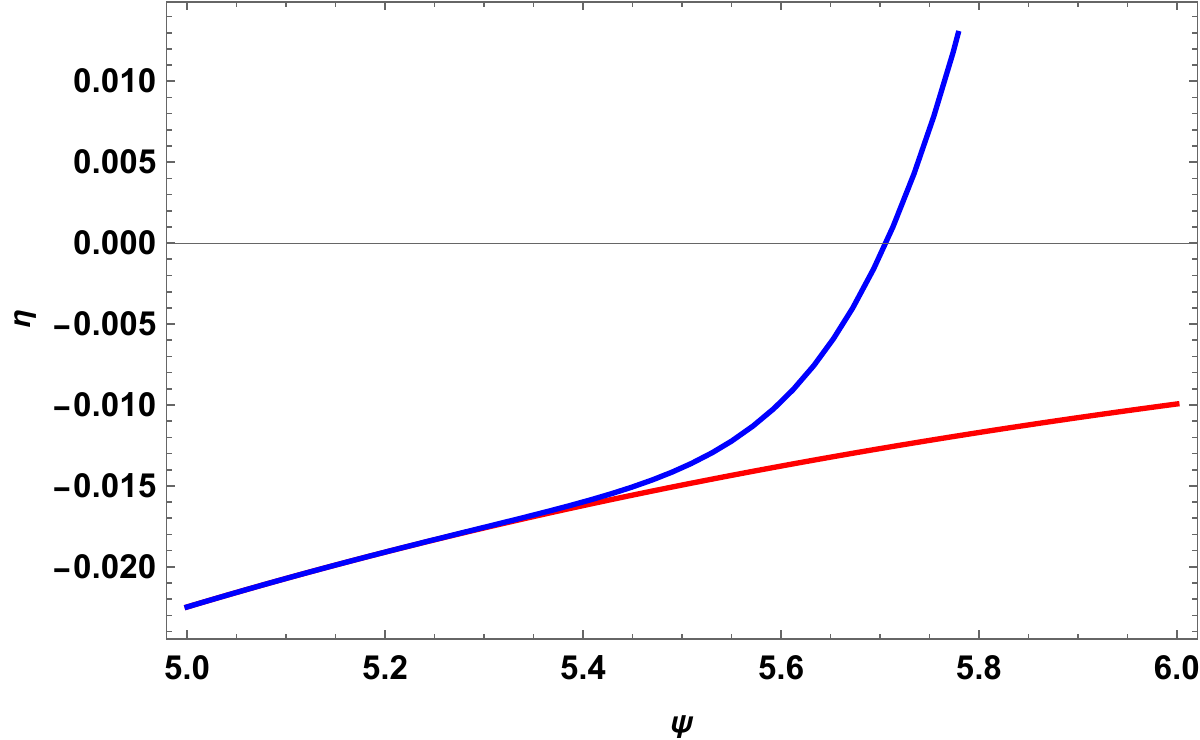}
    \caption{(Left) The potential with (blue curve) and without (red curve) Genesis corrections. (Right) The second slow-roll parameter with (blue curve) and without (red curve) genesis corrections. All plots are generated using the parameter choice given in eq.~\eqref{parameters}.}
    \label{fig: Potential and eta for psi}
\end{figure}

Next, based on the Starobinsky potential with Genesis corrections, eq.~\eqref{eq: psi model}, one can provide qualitative arguments explaining why the scalar power spectrum tilt increases. Specifically, when the corrections to the Starobinsky model are small, the spectral index $n_s$ is primarily determined by the second slow-roll parameter, $\eta_V = \frac{V^{\prime\prime}}{V}$.
Consequently, higher values of $\eta_V$ correspond to larger values of $n_s$. In figure~\ref{fig: Potential and eta for psi}, we plot both the pure Starobinsky potential and the second slow-roll parameter $\eta_V$ for typical values of the field $\psi$. Thus, these qualitative arguments are in full agreement with the numerical simulations.

It is also noteworthy that the increase or decrease of the spectral index $n_s$ is connected to the sign of the ratio $g/c$. This ratio determines the characteristic value of the Hubble parameter during the Genesis phase, eq.~\eqref{eq: sol Early Genesis HN}. A positive value of $g/c$ leads to an increase in $n_s$, whereas a negative value of $g/c$ results in a decrease of $n_s$ compared to the pure Starobinsky model. However, negative values of $g/c$ are \textit{impossible} in the Genesis model, as the Hubble parameter must remain positive during the early Genesis stage. This requirement ensures that Genesis corresponds to a relatively slow expansion from an asymptotically flat spacetime.

\begin{figure}
    \centering
    \includegraphics[width=0.8\linewidth]{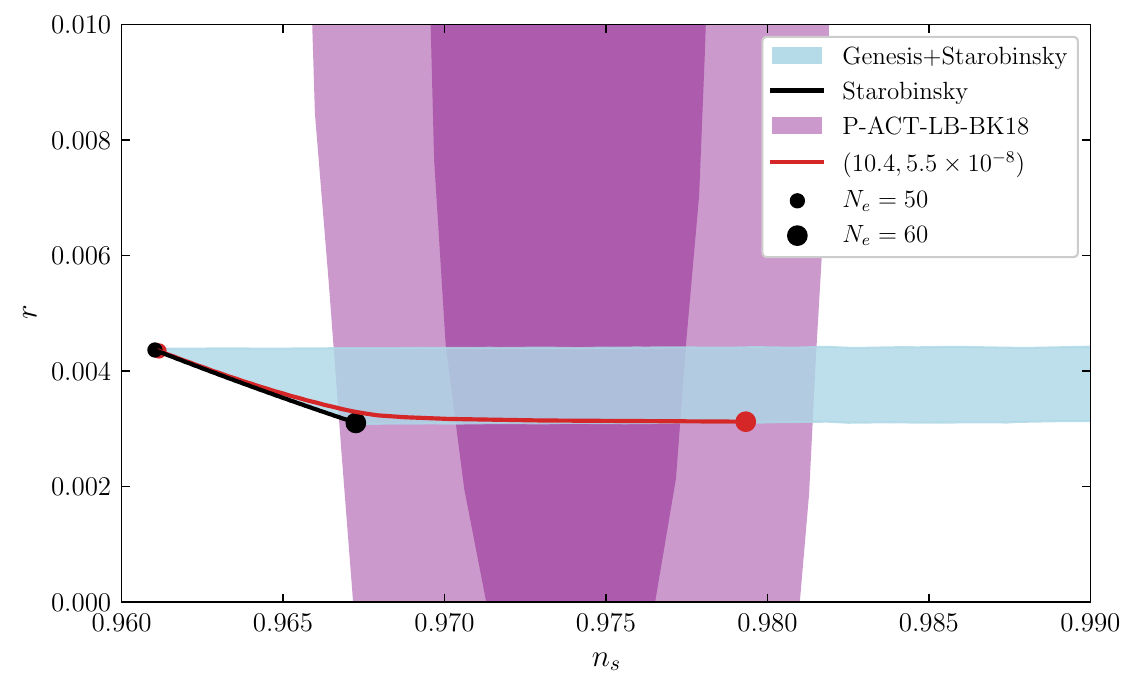}
    \caption{
    Region in the $n_s$-$r$ plane allowed by our model (shaded in blue). The regions corresponding to the constraints from recent observations~\cite{ACT:2025tim} are also shown (shaded in light purple for $1\sigma$ and dark purple for $2\sigma$). The prediction from Starobinsky inflation is shown as a black curve, with small and large dots at the ends representing $N_e=50$ and $N_e=60$, respectively. An example of our model with $n=10.4$ and $g=5.5\times 10^{-8}$ is plotted in red, with small and large dots on the red curve also corresponding to $N_e=50$ and $N_e=60$, respectively.
    }
    \label{fig:experiment}
\end{figure}

Finally, in figure~\ref{fig:experiment}, we show the region in the $n_s$-$r$ plane allowed by our model(shaded in light blue), together with the constraints from recent CMB observations~\cite{ACT:2025tim} (shaded in purple). The light blue region corresponds to the parameter range
\begin{equation}
    9<n<12~,\quad 5\times 10^{-8}<g<10^{-6}~.
\end{equation}
We also extract the values of $n_s$ and $r$ within the range $50<N_e <60$, only when the slow-roll conditions, $\epsilon_H,|\eta_H|<0.1$, are satisfied. As shown in the figure, our model can account for the observed values of $n_s$ and $r$ within the current observational constraints, whereas Starobinsky inflation is disfavored at the $2\sigma$ level. Therefore, our model---where Genesis served as the initial condition for Starobinsky inflation---offers a simple and elegant solution to the tension between  observations and the standard Starobinsky model. 

\section{Conclusion}
\label{sec: Conclusion}

In this article, we have proposed a novel non-singular cosmological scenario within Horndeski gravity, in which the Universe emerges from an asymptotically flat Genesis phase and smoothly transitions, over $\mathcal{O}(1 \textrm{--} 3)$ e-folds, into Starobinsky inflation.~\footnote{Our scenario differs crucially from earlier Genesis models, such as Ref.~\cite{GilChoi:2025hbs}, by ending in a Starobinsky inflationary epoch rather than kination.} This construction is fully consistent within a viable parameter space: the theory remains weakly coupled, is free from ghost and gradient instabilities, and respects luminal (for tensors) and subluminal (for scalars) propagation throughout the entire evolution.

A key outcome of our analysis is that the Genesis phase leaves robust and calculable imprints on the inflationary dynamics. In particular, it generates characteristic corrections to the Starobinsky potential that cannot be captured by simple $\sum_i c_i R^i$-type extensions~\cite{Cheong:2020rao, Rodrigues-da-Silva:2022qiq, Kim:2025dyi}. These corrections consistently enhance the scalar spectral index ($n_s$), thereby improving the agreement of Starobinsky inflation with the latest CMB observations, such as those from ACT~\cite{ACT:2025fju, ACT:2025tim}.

In this work, we have studied a particular subclass of transition functions ensuring theoretical consistency, such as stability, subluminal propagation, and unitarity. Exploring more general realizations of the Genesis-inflation transition, along with their possible signatures in cosmological observables, will be an exciting direction for future research.

\section*{ACKNOWLEDGMENTS}

We are grateful to Masahide Yamaguchi and Hyun Seok Yang for useful discussion and comments. H.G. and P.P. are supported by IBS under the project code, IBS-R018-D3. 
S.C.P. is supported by the National Research Foundation of Korea (NRF) grant funded by the Korea government (MSIT) RS-2023-00283129 and RS-2024-00340153.

\newpage
\appendix

\section{Leading - order expression for the lapse function}
\numberwithin{equation}{section}
\label{App1}

In this appendix, we derive the leading-order correspondence between the value of the lapse function and the field value $\phi$ during the slow-roll Starobinsky inflationary stage. 
From the Friedmann equations with a potential $V(\phi)$, one obtains the following relations:
\begin{equation}\label{a1_1}
    3H^2 = \frac{1}{2}\left(\frac{d\phi}{dt_c}\right)^2+V(\phi)~,
\end{equation}
and
\begin{equation}\label{a1_2}
    \frac{dH}{d\phi} = -\frac{1}{2}\frac{d\phi}{dt_c}~.
\end{equation}
If $V(\phi)$ can be decomposed into a flat plateau and a small correction term as
\begin{equation*}
    V(\phi) = V_0 + \delta V(\phi),
\end{equation*}
then the background solution for the Hubble parameter can also be decomposed as follows
\begin{equation*}
    H = H_0 +  \delta H(\phi)~.
\end{equation*}

Therefore, one obtains
\begin{equation}
\begin{split}
    H_0 & =\sqrt{\frac{V_0}{3}}~,\\    
    \delta H & = \frac{H_0}{2}\frac{\delta V(\phi)}{V_0} ~,\\
    \frac{d\phi}{dt_c} & =- H_0 \frac{\delta V'(\phi)}{V_0}~.
\end{split}
\end{equation}
If the potential corresponds to the Starobinsky model, then  we have
\begin{equation}
    V(\phi) = \frac{3}{4}M_0^2\left(1 - e^{-\sqrt{\frac{2}{3}}\phi}\right)^2 \simeq \frac{3}{4}M_0^2 - \frac{3}{2}M_0^2 e^{-\sqrt{\frac{2}{3}}\phi}.
\end{equation}
Here, we assume that the field $\phi$ is sufficiently large in comparison with the Planck mass. Thus, one arrives at
\begin{equation}
\begin{split}
    H_0 &= \frac{M_0}{2}\\
    \delta H &=  -\frac{1}{2}M_0 e^{-\sqrt{\frac{2}{3}}\phi}\\
    \frac{d\phi}{dt_c} &= -\sqrt{\frac{2}{3}}M_0 e^{-\sqrt{\frac{2}{3}}\phi}
\end{split}
\end{equation}

Using the above results, we find that for Starobinsky inflation, the connection between the lapse function and the field value is as follows:   
\begin{equation*}
    N  = \frac{dt_c}{dt} = \left(\frac{d\phi}{dt_c}\right)^{-1}\left(\frac{d\phi}{dt}\right),
\end{equation*}
\begin{equation}
    N = \frac{c}{M_0}e^{-n+\frac{5}{\sqrt{6}}\phi}~.
\end{equation}
where $t$ is a ADM time coordinate, and $n$ is a positive constant. Here, we also employ the time-slicing defined in eq.~\eqref{eq: time slicing}. This proves the result quoted in subsection~\ref{subsec: Numerical simulation}.

\section{Genesis correction}
\numberwithin{equation}{section}
\label{App2}

In this appendix, we derive the leading-order Genesis corrections in two forms. The first form corresponds to corrections to the potential, while the second involves corrections in the form of $f(R)$ gravity. 
In large-$t$ limit, the Lagrangian functions behave as
\begin{equation}
    A_4 \xrightarrow[]{t\rightarrow\infty} -\frac{1}{2} + O(e^{-2 s t})
\end{equation}
and
\begin{equation}
    A_2 \xrightarrow[]{t\rightarrow\infty}A_2^S +\delta A_2
\end{equation}
Here, $\delta A_2$ is the correction arising from the pre-inflationary Genesis.  In the large $t$ limit the corrections to  $A_4$ are suppressed by the small exponential factor. Therefore, we have $F=\int dX B_{4\phi}/X\simeq 0$.  From the transition formulas between the ADM and covariant formalisms \eqref{FromADMToCov}, one obtains the following expressions for covariant functions:
$$G_2 = A_2 - 2 X F_\phi \simeq A_2$$
and 
$$G_3 = - 2 X F_X- F \simeq, 0$$
one can see that corrections in $G_2 \simeq A_2$ are relevant, i.e, they are less suppressed than the corrections to $G_{3,4}$. The correction $\delta A_2$ can be decomposed into 
\begin{equation}
    \delta A_2 = \delta A_2^{GX} + \delta A_2^{GX^2}  + \delta A_2^{\tau} + \delta  A_2^{e} + O(e^{-2st})~.
\end{equation}
Each term reads
\begin{equation*}
    \delta A_{2}^{GX} = -\frac{1}{2}\frac{g}{N^2}e^{-s t}~,   
\end{equation*}
\begin{equation*}
    \delta A_{2}^{GX^2} = \frac{1}{2}\frac{g}{6N^4}e^{-s t}~.
\end{equation*}

The terms $A_2^{GX,\;GX^2}$ originate from the pre-inflationary Genesis Lagrangian, while  term
$$    \delta A_2^{e} = -A_2^S e^{-s t} + O(e^{-2 s t})
$$
arises from the fact that the entire Starobinsky Lagrangian is coupled to the function $U(t) = (1 + e^{-st})^{-1}$. Finally, the term
$$ 
 \delta A_2^{\tau} = \delta A_2^{\tau_1} + O(1/(ct)^2)
$$
arises  from the modification of Starobinsky Lagrangian due to the replacement of the time variable, namely: $\tau(t) \xrightarrow[]{t\rightarrow\infty}t+c/2$.  Here $ A_2^{\tau_1}$ is given by:
\begin{equation}
    A_2^{\tau_1} = 2e^{-2 n/3}M_0^2 (c t)^{-1/3} -2e^{-4 n/3}M_0^2 (c t)^{1/3} -\frac{4}{3cN^2t^3} ~.
\end{equation}

These functions can be transformed into their covariant form by imposing the following time-slicing:
\begin{equation}
    t= \frac{1}{c}e^{n-\sqrt{\frac{3}{2}}\phi }~,
\end{equation}
\begin{equation}
    N= \frac{c}{\sqrt{3 X}} e^{-n+\sqrt{\frac{3}{2}}\phi}~.
\end{equation}

We find that $\delta A_2^e, \;\delta A_{2}^{GX^2}$ are negligible within $50<N_e<60$. By retaining only $A_2^{GX}$ and $A_2^{\tau_1}$, one can derive the corresponding corrections to the potential as well as equivalent $f(R)$ Lagrangian. As a first step, we assume that the corrections from Genesis scenario are small and expand our Lagrangian with respect to this small parameter. In other words we assume that the combination $ e^{-n + \sqrt{3/2}\phi}$ is sufficiently small to suppress the Genesis corrections. We then canonically normalize the kinetic term with the following field redefinition:
\begin{align*}
    \phi = \frac{e^{-\frac{s e^{n-\sqrt{\frac{3}{2}} \psi
   }}{c}-n-\sqrt{\frac{3}{2}} \psi } \left(3 c g
   e^{n+\sqrt{\frac{3}{2}} \psi }+8 c s^2 e^{\frac{s
   e^{n-\sqrt{\frac{3}{2}} \psi }}{c}+\sqrt{6} \psi }+3 g e^{2 n}
   s\right)}{2 \sqrt{6} c s^2}+\psi\;,
\end{align*}
which leads to a Lagrangian describing Starobinsky inflation together with the leading Genesis corrections:
\begin{align*}
    L_{\psi} &= X_{\psi} - V_{\psi}(\psi)\,,\\
     V_{\psi}(\psi) &= \frac{3}{4} M_0^2 e^{-2 \sqrt{\frac{2}{3}} \psi }
   \left(e^{\sqrt{\frac{2}{3}} \psi }-1\right)^2 + \frac{3 g M_0^2 \left(e^{\sqrt{\frac{2}{3}} \psi }-1\right)
   e^{-\frac{s e^{n-\sqrt{\frac{3}{2}} \psi }}{c}-\frac{7 \psi
   }{\sqrt{6}}} \left(c e^{\sqrt{\frac{3}{2}} \psi }+e^n s\right)}{4
   c s^2} \nonumber
\end{align*}
where $X_{\psi} \equiv -\tfrac{1}{2}\partial_{\mu}\psi \partial^{\mu}\psi$.
Finally, it is straightforward to rewrite this approximate inflationary model in terms of $f(R)$ gravity. 
Proceeding with the well-known redefinitions (for a review, see: \cite{DeFelice:2010aj}) :
\begin{align*}
  \psi(R) = \sqrt{\frac{3}{2}} \text{ln}(f^{\prime})\;,\;\;
  V_{\psi}[\psi(R)] = \frac{f^{\prime} R - f}{2 (f^{\prime})^2}\;,\;\;f \equiv f(R)\;.
\end{align*}
one arrives to:
\begin{align*}
    L_{R} &= f(R),\;\\
    f(R) &=  \frac{R}{2} + \frac{R^2}{12
   M_0^2} 
   \\
   &+ \frac{e^{-\frac{3 \sqrt{3} e^n s}{c
   \left(\frac{R}{M_0^2}+3\right){}^{3/2}}} }{\left(3 M_0^2+R\right){}^2}
   \left(-\frac{3 g M_0^4
   e^n R \sqrt{\frac{3 R}{M_0^2}+9}}{4 c s}-\frac{3 g M_0^2 R^2}{2
   s^2}-\frac{9 g M_0^4 R}{4 s^2}-\frac{g R^3}{4
   s^2}\right)\;.
\end{align*}

\begin{figure}
    \centering
    \includegraphics[width=0.45\linewidth]{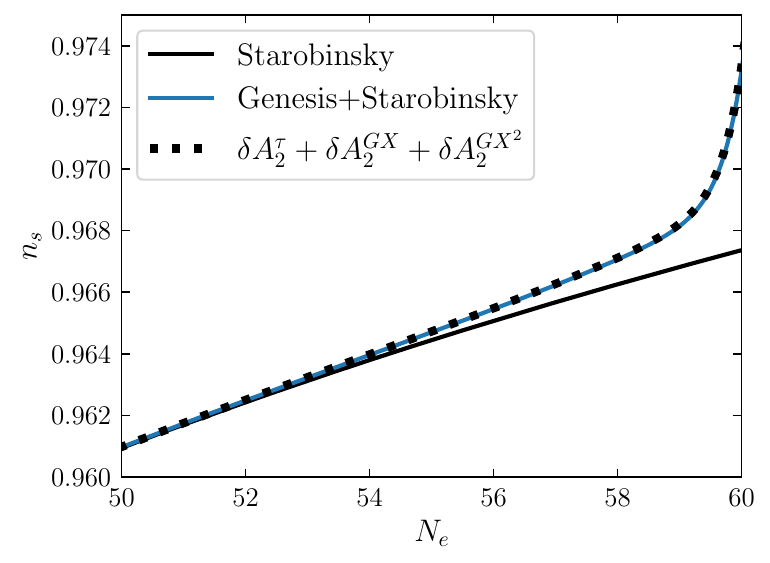}
\includegraphics[width=0.45\linewidth]{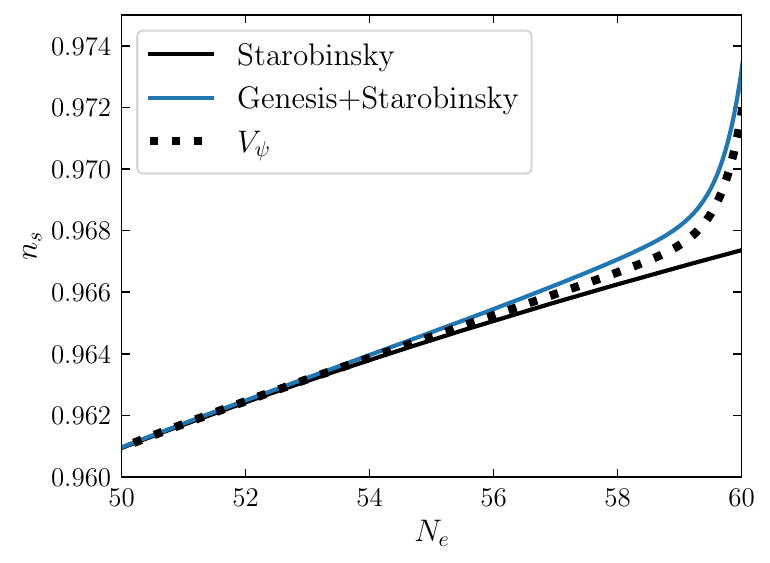}
    \caption{Comparison of $n_s$ values from different models. In the left panel, we show the results from the Starobinsky model, the full model, and the Starobinsky model including the corrections $\delta A_2^{\tau} + \delta A_2^{GX} + \delta A_2^{GX^2}$. In the right panel, we compare the $\psi$ model ($V_\psi$) with the other models. The same parameter values as in figure~\ref{fig:evolution} are used.}
    \label{fig:errors}
\end{figure}

We compute the values of $n_s$ for the Starobinsky model, the full model, and an approximate model that includes the corrections $\delta A_2^{\tau} + \delta A_2^{GX} + \delta A_2^{GX^2}$. In addition, we compare the predictions of the $\psi$ model with those of the Starobinsky and full models. As expected, the differences are sufficiently small, indicating that our approximations are valid. These results are presented in figure~\ref{fig:errors}. We further find that the $\psi$ model captures the leading-order Genesis corrections to the spectral index $n_s$ to a good approximation.

\newpage

\bibliographystyle{JHEPmod}
\bibliography{ref}

\providecommand{\href}[2]{#2}\begingroup\raggedright\begin{thebibliography}{10}

\bibitem{Borde:1996pt}
A.~Borde and A.~Vilenkin, {\it {Singularities in inflationary cosmology: A Review}}, \href{https://doi.org/10.1142/S0218271896000497}{Int. J. Mod. Phys. D {\bfseries 5} (1996) 813} [\href{http://arxiv.org/abs/gr-qc/9612036}{{\ttfamily arXiv:gr-qc/9612036}}].

\bibitem{Horndeski:1974wa}
G.~W. Horndeski, {\it {Second-order scalar-tensor field equations in a four-dimensional space}}, \href{https://doi.org/10.1007/BF01807638}{Int. J. Theor. Phys. {\bfseries 10} (1974) 363}.

\bibitem{Kobayashi:2016xpl}
T.~Kobayashi, {\it {Generic instabilities of nonsingular cosmologies in Horndeski theory: A no-go theorem}}, \href{https://doi.org/10.1103/PhysRevD.94.043511}{Phys. Rev. D {\bfseries 94} (2016) 043511} [\href{http://arxiv.org/abs/1606.05831}{{\ttfamily arXiv:1606.05831}}].

\bibitem{Libanov:2016kfc}
M.~Libanov, S.~Mironov and V.~Rubakov, {\it {Generalized Galileons: instabilities of bouncing and Genesis cosmologies and modified Genesis}}, \href{https://doi.org/10.1088/1475-7516/2016/08/037}{JCAP {\bfseries 08} (2016) 037} [\href{http://arxiv.org/abs/1605.05992}{{\ttfamily arXiv:1605.05992}}].

\bibitem{Zumalacarregui:2013pma}
M.~Zumalac{\'a}rregui and J.~Garc{\'\i}a-Bellido, {\it {Transforming gravity: from derivative couplings to matter to second-order scalar-tensor theories beyond the Horndeski Lagrangian}}, \href{https://doi.org/10.1103/PhysRevD.89.064046}{Phys. Rev. D {\bfseries 89} (2014) 064046} [\href{http://arxiv.org/abs/1308.4685}{{\ttfamily arXiv:1308.4685}}].

\bibitem{Gleyzes:2014dya}
J.~Gleyzes, D.~Langlois, F.~Piazza and F.~Vernizzi, {\it {Healthy theories beyond Horndeski}}, \href{https://doi.org/10.1103/PhysRevLett.114.211101}{Phys. Rev. Lett. {\bfseries 114} (2015) 211101} [\href{http://arxiv.org/abs/1404.6495}{{\ttfamily arXiv:1404.6495}}].

\bibitem{Langlois:2015cwa}
D.~Langlois and K.~Noui, {\it {Degenerate higher derivative theories beyond Horndeski: evading the Ostrogradski instability}}, \href{https://doi.org/10.1088/1475-7516/2016/02/034}{JCAP {\bfseries 02} (2016) 034} [\href{http://arxiv.org/abs/1510.06930}{{\ttfamily arXiv:1510.06930}}].

\bibitem{Ageeva:2021yik}
Y.~Ageeva, P.~Petrov and V.~Rubakov, {\it {Nonsingular cosmological models with strong gravity in the past}}, \href{https://doi.org/10.1103/PhysRevD.104.063530}{Phys. Rev. D {\bfseries 104} (2021) 063530} [\href{http://arxiv.org/abs/2104.13412}{{\ttfamily arXiv:2104.13412}}].

\bibitem{Ageeva:2022asq}
Y.~Ageeva, P.~Petrov and V.~Rubakov, {\it {Generating cosmological perturbations in non-singular Horndeski cosmologies}}, \href{https://doi.org/10.1007/JHEP01(2023)026}{JHEP {\bfseries 01} (2023) 026} [\href{http://arxiv.org/abs/2207.04071}{{\ttfamily arXiv:2207.04071}}].

\bibitem{Akama:2022usl}
S.~Akama and S.~Hirano, {\it {Primordial non-Gaussianity from Galilean Genesis without strong coupling problem}}, \href{https://doi.org/10.1103/PhysRevD.107.063504}{Phys. Rev. D {\bfseries 107} (2023) 063504} [\href{http://arxiv.org/abs/2211.00388}{{\ttfamily arXiv:2211.00388}}].

\bibitem{GilChoi:2025hbs}
H.~Gil~Choi, P.~Petrov and M.~Yamaguchi, {\it {Can Horndeski Genesis be Nonpathological?}},  \href{http://arxiv.org/abs/2503.02626}{{\ttfamily arXiv:2503.02626}}.

\bibitem{Planck:2018nkj}
{\scshape Planck} collaboration, N.~Aghanim et~al., {\it {Planck 2018 results. I. Overview and the cosmological legacy of Planck}}, \href{https://doi.org/10.1051/0004-6361/201833880}{Astron. Astrophys. {\bfseries 641} (2020) A1} [\href{http://arxiv.org/abs/1807.06205}{{\ttfamily arXiv:1807.06205}}].

\bibitem{ACT:2025fju}
{\scshape ACT} collaboration, T.~Louis et~al., {\it {The Atacama Cosmology Telescope: DR6 Power Spectra, Likelihoods and $\Lambda$CDM Parameters}},  \href{http://arxiv.org/abs/2503.14452}{{\ttfamily arXiv:2503.14452}}.

\bibitem{ACT:2025tim}
{\scshape ACT} collaboration, E.~Calabrese et~al., {\it {The Atacama Cosmology Telescope: DR6 Constraints on Extended Cosmological Models}},  \href{http://arxiv.org/abs/2503.14454}{{\ttfamily arXiv:2503.14454}}.

\bibitem{Starobinsky:1980te}
A.~A. Starobinsky, {\it {A New Type of Isotropic Cosmological Models Without Singularity}}, \href{https://doi.org/10.1016/0370-2693(80)90670-X}{Phys. Lett. B {\bfseries 91} (1980) 99}.

\bibitem{Martin:2013tda}
J.~Martin, C.~Ringeval and V.~Vennin, {\it {Encyclop{\ae}dia Inflationaris}: {Opiparous Edition}}, \href{https://doi.org/10.1016/j.dark.2024.101653}{Phys. Dark Univ. {\bfseries 5-6} (2014) 75} [\href{http://arxiv.org/abs/1303.3787}{{\ttfamily arXiv:1303.3787}}].

\bibitem{Ketov:2025nkr}
S.~V. Ketov, {\it {On Legacy of Starobinsky Inflation}},  1, 2025, \href{http://arxiv.org/abs/2501.06451}{{\ttfamily arXiv:2501.06451}}.

\bibitem{Bezrukov:2007ep}
F.~L. Bezrukov and M.~Shaposhnikov, {\it {The Standard Model Higgs boson as the inflaton}}, \href{https://doi.org/10.1016/j.physletb.2007.11.072}{Phys. Lett. B {\bfseries 659} (2008) 703} [\href{http://arxiv.org/abs/0710.3755}{{\ttfamily arXiv:0710.3755}}].

\bibitem{Futamase:1987ua}
T.~Futamase and K.-i. Maeda, {\it {Chaotic Inflationary Scenario in Models Having Nonminimal Coupling With Curvature}}, \href{https://doi.org/10.1103/PhysRevD.39.399}{Phys. Rev. D {\bfseries 39} (1989) 399}.

\bibitem{Park:2008hz}
S.~C. Park and S.~Yamaguchi, {\it {Inflation by non-minimal coupling}}, \href{https://doi.org/10.1088/1475-7516/2008/08/009}{JCAP {\bfseries 08} (2008) 009} [\href{http://arxiv.org/abs/0801.1722}{{\ttfamily arXiv:0801.1722}}].

\bibitem{Hyun:2022uzc}
S.~C. Hyun, J.~Kim, S.~C. Park and T.~Takahashi, {\it {Non-minimally assisted chaotic inflation}}, \href{https://doi.org/10.1088/1475-7516/2022/05/045}{JCAP {\bfseries 05} (2022) 045} [\href{http://arxiv.org/abs/2203.09201}{{\ttfamily arXiv:2203.09201}}].

\bibitem{Hyun:2023bkf}
S.~C. Hyun, J.~Kim, T.~Kodama, S.~C. Park and T.~Takahashi, {\it {Nonminimally assisted inflation: a~general analysis}}, \href{https://doi.org/10.1088/1475-7516/2023/05/050}{JCAP {\bfseries 05} (2023) 050} [\href{http://arxiv.org/abs/2302.05866}{{\ttfamily arXiv:2302.05866}}].

\bibitem{Hamada:2014iga}
Y.~Hamada, H.~Kawai, K.-y. Oda and S.~C. Park, {\it {Higgs Inflation is Still Alive after the Results from BICEP2}}, \href{https://doi.org/10.1103/PhysRevLett.112.241301}{Phys. Rev. Lett. {\bfseries 112} (2014) 241301} [\href{http://arxiv.org/abs/1403.5043}{{\ttfamily arXiv:1403.5043}}].

\bibitem{Hamada:2014wna}
Y.~Hamada, H.~Kawai, K.-y. Oda and S.~C. Park, {\it {Higgs inflation from Standard Model criticality}}, \href{https://doi.org/10.1103/PhysRevD.91.053008}{Phys. Rev. D {\bfseries 91} (2015) 053008} [\href{http://arxiv.org/abs/1408.4864}{{\ttfamily arXiv:1408.4864}}].

\bibitem{He:2018mgb}
M.~He, R.~Jinno, K.~Kamada, S.~C. Park, A.~A. Starobinsky and J.~Yokoyama, {\it {On the violent preheating in the mixed Higgs-$R^2$ inflationary model}}, \href{https://doi.org/10.1016/j.physletb.2019.02.008}{Phys. Lett. B {\bfseries 791} (2019) 36} [\href{http://arxiv.org/abs/1812.10099}{{\ttfamily arXiv:1812.10099}}].

\bibitem{Cheong:2022gfc}
D.~Y. Cheong, K.~Kohri and S.~C. Park, {\it {The inflaton that could: primordial black holes and second order gravitational waves from tachyonic instability induced in Higgs-R $^{2}$ inflation}}, \href{https://doi.org/10.1088/1475-7516/2022/10/015}{JCAP {\bfseries 10} (2022) 015} [\href{http://arxiv.org/abs/2205.14813}{{\ttfamily arXiv:2205.14813}}].

\bibitem{Cheong:2019vzl}
D.~Y. Cheong, S.~M. Lee and S.~C. Park, {\it {Primordial black holes in Higgs-$R^2$ inflation as the whole of dark matter}}, \href{https://doi.org/10.1088/1475-7516/2021/01/032}{JCAP {\bfseries 01} (2021) 032} [\href{http://arxiv.org/abs/1912.12032}{{\ttfamily arXiv:1912.12032}}].

\bibitem{Ema:2017rqn}
Y.~Ema, {\it {Higgs Scalaron Mixed Inflation}}, \href{https://doi.org/10.1016/j.physletb.2017.04.060}{Phys. Lett. B {\bfseries 770} (2017) 403} [\href{http://arxiv.org/abs/1701.07665}{{\ttfamily arXiv:1701.07665}}].

\bibitem{Jinno:2019und}
R.~Jinno, M.~Kubota, K.-y. Oda and S.~C. Park, {\it {Higgs inflation in metric and Palatini formalisms: Required suppression of higher dimensional operators}}, \href{https://doi.org/10.1088/1475-7516/2020/03/063}{JCAP {\bfseries 03} (2020) 063} [\href{http://arxiv.org/abs/1904.05699}{{\ttfamily arXiv:1904.05699}}].

\bibitem{Koh:2023zgn}
S.~Koh, S.~C. Park and G.~Tumurtushaa, {\it {Higgs inflation with a Gauss-Bonnet term}}, \href{https://doi.org/10.1103/PhysRevD.110.023523}{Phys. Rev. D {\bfseries 110} (2024) 023523} [\href{http://arxiv.org/abs/2308.00897}{{\ttfamily arXiv:2308.00897}}].

\bibitem{Cheong:2018udx}
D.~Y. Cheong, S.~M. Lee and S.~C. Park, {\it {Higgs Inflation and the Refined dS Conjecture}}, \href{https://doi.org/10.1016/j.physletb.2018.12.046}{Phys. Lett. B {\bfseries 789} (2019) 336} [\href{http://arxiv.org/abs/1811.03622}{{\ttfamily arXiv:1811.03622}}].

\bibitem{Lee:2020yaj}
S.~M. Lee, K.-y. Oda and S.~C. Park, {\it {Spontaneous Leptogenesis in Higgs Inflation}}, \href{https://doi.org/10.1007/JHEP03(2021)083}{JHEP {\bfseries 03} (2021) 083} [\href{http://arxiv.org/abs/2010.07563}{{\ttfamily arXiv:2010.07563}}].

\bibitem{Park:2024ceu}
S.~C. Park, {\it {Non-minimally coupled quintessential inflation}},  \href{http://arxiv.org/abs/2412.08833}{{\ttfamily arXiv:2412.08833}}.

\bibitem{Cheong:2021vdb}
D.~Y. Cheong, S.~M. Lee and S.~C. Park, {\it {Progress in Higgs inflation}}, \href{https://doi.org/10.1007/s40042-021-00086-2}{J. Korean Phys. Soc. {\bfseries 78} (2021) 897} [\href{http://arxiv.org/abs/2103.00177}{{\ttfamily arXiv:2103.00177}}].

\bibitem{Planck:2018jri}
{\scshape Planck} collaboration, Y.~Akrami et~al., {\it {Planck 2018 results. X. Constraints on inflation}}, \href{https://doi.org/10.1051/0004-6361/201833887}{Astron. Astrophys. {\bfseries 641} (2020) A10} [\href{http://arxiv.org/abs/1807.06211}{{\ttfamily arXiv:1807.06211}}].

\bibitem{BICEP:2021xfz}
{\scshape BICEP, Keck} collaboration, P.~A.~R. Ade et~al., {\it {Improved Constraints on Primordial Gravitational Waves using Planck, WMAP, and BICEP/Keck Observations through the 2018 Observing Season}}, \href{https://doi.org/10.1103/PhysRevLett.127.151301}{Phys. Rev. Lett. {\bfseries 127} (2021) 151301} [\href{http://arxiv.org/abs/2110.00483}{{\ttfamily arXiv:2110.00483}}].

\bibitem{Penrose:1964wq}
R.~Penrose, {\it {Gravitational collapse and space-time singularities}}, \href{https://doi.org/10.1103/PhysRevLett.14.57}{Phys. Rev. Lett. {\bfseries 14} (1965) 57}.

\bibitem{Borde:2001nh}
A.~Borde, A.~H. Guth and A.~Vilenkin, {\it {Inflationary space-times are incompletein past directions}}, \href{https://doi.org/10.1103/PhysRevLett.90.151301}{Phys. Rev. Lett. {\bfseries 90} (2003) 151301} [\href{http://arxiv.org/abs/gr-qc/0110012}{{\ttfamily arXiv:gr-qc/0110012}}].

\bibitem{Lesnefsky:2022fen}
J.~E. Lesnefsky, D.~A. Easson and P.~C.~W. Davies, {\it {Past-completeness of inflationary spacetimes}}, \href{https://doi.org/10.1103/PhysRevD.107.044024}{Phys. Rev. D {\bfseries 107} (2023) 044024} [\href{http://arxiv.org/abs/2207.00955}{{\ttfamily arXiv:2207.00955}}].

\bibitem{Arkani-Hamed:2006emk}
N.~Arkani-Hamed, L.~Motl, A.~Nicolis and C.~Vafa, {\it {The String landscape, black holes and gravity as the weakest force}}, \href{https://doi.org/10.1088/1126-6708/2007/06/060}{JHEP {\bfseries 06} (2007) 060} [\href{http://arxiv.org/abs/hep-th/0601001}{{\ttfamily arXiv:hep-th/0601001}}].

\bibitem{Kobayashi:2011nu}
T.~Kobayashi, M.~Yamaguchi and J.~Yokoyama, {\it {Generalized G-inflation: Inflation with the most general second-order field equations}}, \href{https://doi.org/10.1143/PTP.126.511}{Prog. Theor. Phys. {\bfseries 126} (2011) 511} [\href{http://arxiv.org/abs/1105.5723}{{\ttfamily arXiv:1105.5723}}].

\bibitem{Deffayet:2010qz}
C.~Deffayet, O.~Pujolas, I.~Sawicki and A.~Vikman, {\it {Imperfect Dark Energy from Kinetic Gravity Braiding}}, \href{https://doi.org/10.1088/1475-7516/2010/10/026}{JCAP {\bfseries 10} (2010) 026} [\href{http://arxiv.org/abs/1008.0048}{{\ttfamily arXiv:1008.0048}}].

\bibitem{Kobayashi:2019hrl}
T.~Kobayashi, {\it {Horndeski theory and beyond: a review}}, \href{https://doi.org/10.1088/1361-6633/ab2429}{Rept. Prog. Phys. {\bfseries 82} (2019) 086901} [\href{http://arxiv.org/abs/1901.07183}{{\ttfamily arXiv:1901.07183}}].

\bibitem{DeFelice:2014bma}
A.~De~Felice and S.~Tsujikawa, {\it {Inflationary gravitational waves in the effective field theory of modified gravity}}, \href{https://doi.org/10.1103/PhysRevD.91.103506}{Phys. Rev. D {\bfseries 91} (2015) 103506} [\href{http://arxiv.org/abs/1411.0736}{{\ttfamily arXiv:1411.0736}}].

\bibitem{Kobayashi:2015gga}
T.~Kobayashi, M.~Yamaguchi and J.~Yokoyama, {\it {Galilean Creation of the Inflationary Universe}}, \href{https://doi.org/10.1088/1475-7516/2015/07/017}{JCAP {\bfseries 07} (2015) 017} [\href{http://arxiv.org/abs/1504.05710}{{\ttfamily arXiv:1504.05710}}].

\bibitem{Ageeva:2018lko}
Y.~A. Ageeva, O.~A. Evseev, O.~I. Melichev and V.~A. Rubakov, {\it {Horndeski Genesis: strong coupling and absence thereof}}, \href{https://doi.org/10.1051/epjconf/201819107010}{EPJ Web Conf. {\bfseries 191} (2018) 07010} [\href{http://arxiv.org/abs/1810.00465}{{\ttfamily arXiv:1810.00465}}].

\bibitem{Ageeva:2020gti}
Y.~Ageeva, O.~Evseev, O.~Melichev and V.~Rubakov, {\it {Toward evading the strong coupling problem in Horndeski genesis}}, \href{https://doi.org/10.1103/PhysRevD.102.023519}{Phys. Rev. D {\bfseries 102} (2020) 023519} [\href{http://arxiv.org/abs/2003.01202}{{\ttfamily arXiv:2003.01202}}].

\bibitem{Coule:1987wt}
D.~H. Coule and M.~B. Mijic, {\it {Quantum Fluctuations and Eternal Inflation in the $r^{2}$ Model}}, \href{https://doi.org/10.1142/S0217751X88000266}{Int. J. Mod. Phys. A {\bfseries 3} (1988) 617}.

\bibitem{Maeda:1987xf}
K.-i. Maeda, {\it {Inflation as a Transient Attractor in R**2 Cosmology}}, \href{https://doi.org/10.1103/PhysRevD.37.858}{Phys. Rev. D {\bfseries 37} (1988) 858}.

\bibitem{Starobinsky:1983zz}
A.~A. Starobinsky, {\it {The Perturbation Spectrum Evolving from a Nonsingular Initially De-Sitter Cosmology and the Microwave Background Anisotropy}}, {Sov. Astron. Lett. {\bfseries 9} (1983) 302}.

\bibitem{Maldacena:2002vr}
J.~M. Maldacena, {\it {Non-Gaussian features of primordial fluctuations in single field inflationary models}}, \href{https://doi.org/10.1088/1126-6708/2003/05/013}{JHEP {\bfseries 05} (2003) 013} [\href{http://arxiv.org/abs/astro-ph/0210603}{{\ttfamily arXiv:astro-ph/0210603}}].

\bibitem{Drees:2025ngb}
M.~Drees and Y.~Xu, {\it {Refined predictions for Starobinsky inflation and post-inflationary constraints in light of ACT}}, \href{https://doi.org/10.1016/j.physletb.2025.139612}{Phys. Lett. B {\bfseries 867} (2025) 139612} [\href{http://arxiv.org/abs/2504.20757}{{\ttfamily arXiv:2504.20757}}].

\bibitem{Addazi:2025qra}
A.~Addazi, Y.~Aldabergenov and S.~V. Ketov, {\it {Curvature corrections to Starobinsky inflation can explain the ACT results}},  \href{http://arxiv.org/abs/2505.10305}{{\ttfamily arXiv:2505.10305}}.

\bibitem{Dioguardi:2025mpp}
C.~Dioguardi and A.~Karam, {\it {Palatini linear attractors are back in action}}, \href{https://doi.org/10.1103/23b3-9d7q}{Phys. Rev. D {\bfseries 111} (2025) 123521} [\href{http://arxiv.org/abs/2504.12937}{{\ttfamily arXiv:2504.12937}}].

\bibitem{He:2025bli}
M.~He, M.~Hong and K.~Mukaida, {\it {Increase of $n_s$ in regularized pole inflation {\&} Einstein-Cartan gravity}},  \href{http://arxiv.org/abs/2504.16069}{{\ttfamily arXiv:2504.16069}}.

\bibitem{DeFelice:2011zh}
A.~De~Felice and S.~Tsujikawa, {\it {Primordial non-Gaussianities in general modified gravitational models of inflation}}, \href{https://doi.org/10.1088/1475-7516/2011/04/029}{JCAP {\bfseries 04} (2011) 029} [\href{http://arxiv.org/abs/1103.1172}{{\ttfamily arXiv:1103.1172}}].

\bibitem{Cheong:2020rao}
D.~Y. Cheong, H.~M. Lee and S.~C. Park, {\it {Beyond the Starobinsky model for inflation}}, \href{https://doi.org/10.1016/j.physletb.2020.135453}{Phys. Lett. B {\bfseries 805} (2020) 135453} [\href{http://arxiv.org/abs/2002.07981}{{\ttfamily arXiv:2002.07981}}].

\bibitem{Rodrigues-da-Silva:2022qiq}
G.~Rodrigues-da Silva and L.~G. Medeiros, {\it {Second-order corrections to Starobinsky inflation}}, \href{https://doi.org/10.1140/epjc/s10052-023-12149-8}{Eur. Phys. J. C {\bfseries 83} (2023) 1032} [\href{http://arxiv.org/abs/2207.02103}{{\ttfamily arXiv:2207.02103}}].

\bibitem{Kim:2025dyi}
J.~Kim, X.~Wang, Y.-l. Zhang and Z.~Ren, {\it {Enhancement of primordial curvature perturbations in $R^3$-corrected Starobinsky-Higgs inflation}},  \href{http://arxiv.org/abs/2504.12035}{{\ttfamily arXiv:2504.12035}}.

\bibitem{DeFelice:2010aj}
A.~De~Felice and S.~Tsujikawa, {\it {f(R) theories}}, \href{https://doi.org/10.12942/lrr-2010-3}{Living Rev. Rel. {\bfseries 13} (2010) 3} [\href{http://arxiv.org/abs/1002.4928}{{\ttfamily arXiv:1002.4928}}].

\end{thebibliography}\endgroup

\end{document}